\documentclass[reprint,aps,prx,twocolumn,superscriptaddress,amsmath,amssymb,floatfix]{revtex4-2}
\usepackage{graphicx}
\usepackage{layout}
\usepackage{dcolumn}
\usepackage{braket}
\usepackage{bm}
\usepackage{color}
\usepackage{verbatim}
\usepackage{hyperref}
\usepackage{ulem}
\usepackage{enumitem}
\usepackage{multirow}

\begin{document}

\title{Nonequilibrium Cooper quartet generation in superconducting devices}

\author{Luca Chirolli}
\affiliation{Department of Physics and Astronomy, University of Florence, I-50019 Sesto Fiorentino, Italy}

\author{Alessandro Braggio}
\affiliation{NEST, Istituto Nanoscienze CNR, I-56127 Pisa, Italy}

\author{Michele Governale}
\affiliation{School of Chemical and Physical Sciences and MacDiarmid Institute for Advanced Materials and Nanotechnology, Victoria University of Wellington, PO Box 600, Wellington 6140, New Zealand}

\begin{abstract}
Cooper quartets are aggregates of four electrons that generalize the concept of Cooper pairs, and their study can unfold unexplored perspectives in correlated matter and many-body physics. We propose a method to isolate them in a double-quantum-dot system coupled to conventional superconducting and normal leads. By driving the system out of equilibrium, we show that a resonance between the vacuum $|0\rangle$ and the four-electron state $|4e\rangle$ emerges in the high bias voltage regime, which involves a two-Cooper pair exchange process and is characterized by finite quartet correlations. We study the transport properties of the system and show that a peak in the Andreev current at high bias voltage has a width that scales with the magnitude of the quartet coupling $\Gamma_{4e}$, which can be tuned by the phase of additional superconducting leads, yielding distinctive signatures. By further studying the current-current correlations and the Fano factor, we establish a regime characterized by equal auto- and cross-correlations, which we interpret as a definitive signature of fast coherent two-Cooper-pair oscillations between the dots and the superconducting leads. The proposed platform, experimentally accessible in a quantum solid-state laboratory, enables exploration of quartet correlations and multifermion-correlated states of matter. 
\end{abstract}

\maketitle

\section{Introduction}
Cooper quartets are the generalization of Cooper pairs underlying superconductivity. Their condensation in a phase-coherent superfluid state may constitute a charge-$4e$ superconductor. However, Cooper quartets become a distinct phase only when they are more stable than pairs. This suggests that stabilizing such an exotic state requires suppressing conventional charge-$2e$ pairing. In the literature, several theoretical studies have explored systems that may host a charge-$4e$ superconducting phase, broadly classified into two categories: fermionic and bosonic. As far as fermionic systems are concerned, a confined many-body quantum phase of tightly-bound Cooper quartets emerges in four-component fermion models at low density and strong interactions \cite{wu2005competing,lecheminant2005,capponi2007confinement,capponi2008molecular,soldini2024charge4e,guan2009unified}, with a confinement-deconfinement transition between a quartet state and a paired state being controlled by a discrete Ising symmetry breaking \cite{lecheminant2005}. Treating Cooper pairs as bosons, a paired-superfluid phase appears in the phase diagram of two-component bosonic systems  \cite{kuklov2004superfluid,kuklov2004twocomponent}, and in general an interference-correlated pair-superfluid phase emerges in multi-component systems featuring flat bands \cite{doucot2002pairing}. In particular, in a rhombi chain showing the Aharonov-Bhom caging phenomenon \cite{vidal1998ABcages}, destructive interference localizes single-particle states, leaving delocalized two-particle states \cite{rizzi20064e-condensation,rizzi2006phase,cartwright2018rhombi-chain,chen2025}. Furthermore, in the weak coupling regime, it has been suggested that preformed Cooper pairs may acquire coherence by pairing up in a two-Cooper pair state, and charge-$4e$ superconductivity has been proposed in the fluctuating state of two-component superconductors \cite{babaev2004superconductor}, in the pair-density wave state \cite{radzihovsky2009,berg2009charge-4e,radzihovsky2011,wu2024d-wave}, in nematic superconductors \cite{fernandes2021charge-4e,jian2021charge-4e,wu2024d-wave}, and in twisted cuprates \cite{liu2023charge-4e}. Higher-order composite states characterized by a generalized quadrupling order have been proposed \cite{maccari2022effects,maccari2023prediction,babaev2024topological} and possibly observed \cite{grinenko2021state}. To date, a charge-$4e$ superconductor has not yet been experimentally demonstrated. However, the study of fermionic quartets \cite{volovik2024fermionic} may offer a deeper understanding in the physics of many-body systems \cite{mizel2004three,peng2009quantum,dai2017four-body,kumar2020large-scale,zhang2022synthesizing}. 

An appealing route to isolate Cooper quartets relies on quantum engineering in hybrid superconducting platforms. 
Several key ideas—based on the interference mechanism proposed in Ref.~\cite{doucot2002pairing}---have been successfully demonstrated, thereby realizing an engineered two-Cooper-pair state in these devices. In superconducting circuits composed of islands connected by Josephson junctions and arranged in a rhombic chain, a two-Cooper pair ground state emerges \cite{blatter2001design,doucot2002pairing}. The latter was proposed as a parity-protected qubit \cite{ioffe2002topologically,ioffe2002possible,doucot2003topological,doucot2005protected,kitaev2006protected,doucot2012physical}, and experimentally realized in networks of Josephson junctions \cite{gladchenko2009superconducting,bell2014protected,bondar2025}, and in hybrid superconducting devices \cite{larsen2020parity-protected,messelot2024phase,arnault2025multiplet,leblanc2025gate,banszerus2024voltage-controlled,ciaccia2024charge-4e,banszerus2025hybrid}. In the context of superconducting devices, a quartet current has been theoretically proposed \cite{cuevas2007voltage-induced,freyn2011production,jonckheere2013multipair} and experimentally realized \cite{pfeffer2014subgap,cohen2018nonlocal,huang2022evidence} in a voltage-biased three-terminal setup. In this system, for a proper choice of the voltage biases of the three terminals, the AC-Josephson effect cancels, resulting in a DC-Josephson current of two Cooper pairs.

With the goal of engineering a correlated Cooper quartet state in hybrid superconducting devices, some of the authors, in Ref.~\cite{chirolli2024}, showed that a system of two quantum dots coupled to superconducting leads can develop finite quartet correlations in the presence of attractive interactions between the dots. The result is based on the observation that quartet correlations naturally arise when a coupling between the vacuum and the fourfold occupied state selects a quartet ground state of the form 
\begin{equation}\label{Eq:EquilibriumGS}
|\psi^\pm_{\rm Q}\rangle=\frac{1}{\sqrt{2}}(|0\rangle\pm |4e\rangle).
\end{equation}
Such a condition can be achieved by bringing the vacuum and the fourfold occupied state in resonance, so that higher order processes, involving exchange of two Cooper pairs with adjacent standard superconducting leads, may mediate a quartet coherent coupling.   However, in equilibrium, this condition requires engineering an attractive interaction in a double-quantum-dot system, which is a challenging task \cite{moser2013,delbecq2013,hamo2016,bhattacharya2021,vigneau2022}.

A possible way to overcome this limitation in a system with naturally repulsive Coulomb interactions is to drive it out of equilibrium. In this work, we explore this idea and study Cooper-quartet correlations in a double quantum dot (DQD) system coupled to conventional superconducting and normal leads, see Fig.~\ref{fig-outline}(a), and brought out of equilibrium by a voltage bias. The latter induces a non-equilibrium redistribution of state populations, which allows us to probe the high-energy part of the spectrum, where, in the presence of repulsive interactions, the quartet subspace resides. Finally, we show that a {\it quartet resonance} appears in the high-bias transport regime and manifests itself in observables such as the Andreev current through the system and the current correlations quantified by the Fano factor. 

The system considered has been widely studied in the context of resonant Andreev tunneling, Josephson effect in the presence of interactions \cite{ishizaka1995,fazio1998,fazio1999erratum,clerk2000,clerk2000andreev,recher2001,recher2003,pala2007,governale2008,hussein2014,sothmann2014unconventional}, and Cooper pair splitting \cite{recher2001,recher2003,sauret2004quantum,chevallier2011current,burset2011microscopic,hussein2016,brange2021dynamic,brange2024adiabatic,delfino2025cooper-pair}, with several successful experimental demonstrations 
\cite{russo2005experimental,hofstetter2009cooper,hofstetter2011finite-bias,schindele2012near-unity,das2012high-efficiency,fulop2014local,tan2015cooper,fulop2015magnetic,borzenets2016high,baba2018cooper-pair,pandey2021ballistic,ranni2021real-time,scherubl2022from,kurtossy2022parallel,bordoloi2022spin,wang2022singlet,dejong2023controllable,bordin2024crossed,wang2023triplet}. Yet, the high-bias regime that we study in the present work hides a quartet subspace that to the best of our knowledge has never been described neither investigated, thus opening a new avenue for the experimental investigation of quartet correlations.

\section{Outline of main results}
\label{Sec:Outline}

\begin{figure*}[t]
	\centering
	\includegraphics[width=2.0\columnwidth]{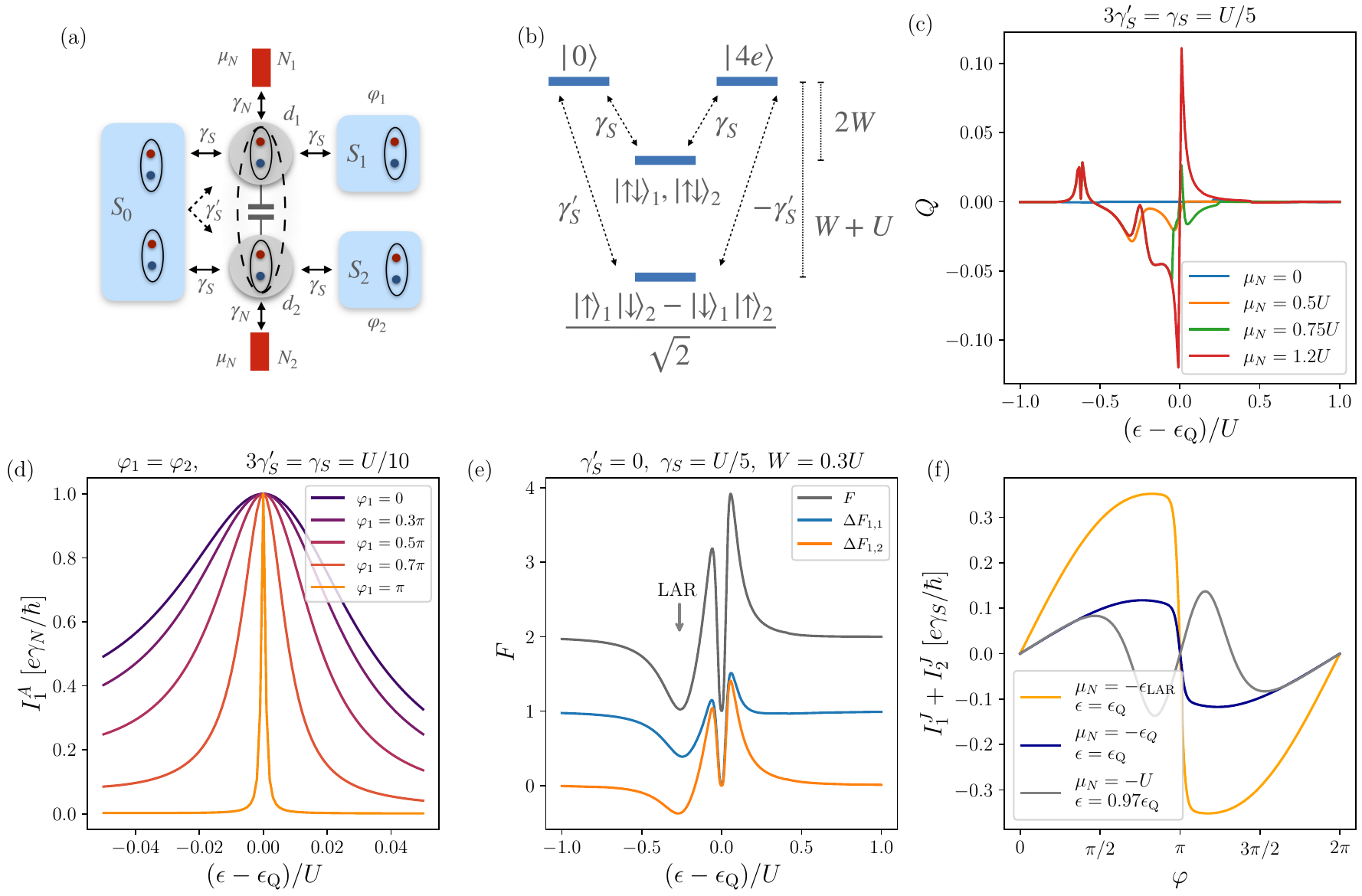}
	\caption{(a) Schematics of the system constituted by two quantum dots coupled to a common superconducting lead, $S_0$, via tunneling rates $\gamma_S$ and $\gamma_S'$, that induce local and non-local proximity pairing terms on the double quantum dot system, and to two normal leads, $N_1$ and $N_2$, at chemical potential $\mu_N$, with respect to the superconducting lead, and coupled via rates $\gamma_N$. We also consider additional superconducting leads, $S_1$ and $S_2$, each coupled to the dots at rate $\gamma_S$, with the convention that by default, when the phases play no role, we refer to the one-SC-terminal setup. (b) Scheme of the effective quartet coupling between the vacuum $|0\rangle$ and the four-electron state $|4e\rangle$, arising from the second order processes involving the two electron states $|d+\rangle=\frac{1}{\sqrt 2}(|\!\uparrow\downarrow\rangle_1+|\!\uparrow\downarrow\rangle_2)$ and $|S\rangle=\frac{1}{\sqrt{2}}(|\!\uparrow\rangle_1|\!\downarrow\rangle_2-|\!\downarrow\rangle_1|\!\uparrow\rangle_2)$ located at energy $-2W$ and $-U-W$ with respect to the zero energy reference of the vacuum state and the four-electron state, at the resonance condition $\epsilon=\epsilon_{\rm Q}$. (c) Quartet correlator $Q$ for one-superconducting-terminal setup as a function of the dot level $\epsilon$, for a few values of the normal leads chemical potential $\mu_N$. (d) Peak in the Andreev current at high bias voltage as a function of the dot level $\epsilon$, for the case of a three-superconducting terminal setup, evaluated for few values of the phase $\varphi_1=\varphi_2$. (e) Fano factor of the sum of the currents in the normal leads at high bias voltage as a function of the dot level $\epsilon$ for the case $\gamma_S'=0$, together with the breakdown $\Delta F_{1,1}$ and $\Delta F_{1,2}$ contributions. (f) Josephson current flowing between $S_0$ and $S_1$ and $S_2$ as a function of the phase $\varphi=\varphi_1=\varphi_2$, for three pairs of values $(\epsilon,\mu_N)$, that are also reported in Fig.~\ref{fig-josephson}(a,b) by dashed lines.}
	\label{fig-outline}
\end{figure*}

The first point we must clarify is the definition of Cooper quartets and the means for generating and detecting them. By analogy with the concept that Cooper pairs are 
characterized by the onset of finite pair correlations, Cooper quartets are characterized by finite quartet correlations: a four-point correlation function of the form \cite{chirolli2024}
\begin{equation}\label{Eq:correlatorQ}
Q=\langle d_{1\downarrow}d_{1\uparrow}d_{2\downarrow}d_{2\uparrow}\rangle-\langle d_{1\downarrow}d_{1\uparrow}\rangle\langle d_{2\downarrow}d_{2\uparrow}\rangle-\langle d_{1\downarrow}d_{2\uparrow}\rangle\langle d_{1\uparrow}d_{2\downarrow}\rangle,
\end{equation}
where $d_{i\sigma}$ are fermionic annihilation operators of spin $\sigma$ and generalized orbital label $i=1,2$. In the present context, the latter refers to two quantum dots. Clearly, the quartet correlator $Q$ acquires a finite value when evaluated on states of the kind of Eq.~\eqref{Eq:EquilibriumGS}. Furthermore, the definition Eq.~\eqref{Eq:correlatorQ} eliminates all the cases where four-electron correlations emerge from two-electron correlations, thus emphasizing the crucial role of the interactions in developing finite quartet correlations.

The system we study is shown in Fig.~\ref{fig-outline}(a) and is fully discussed in Sec.~\ref{Sec:System}. It is composed of two quantum dots coupled to superconducting and normal leads. In the presence of repulsive intradot $U$ and interdot $W$ density-density Coulomb interaction ($U,W>0$), the vacuum and fourfold occupied state become degenerate when the energy level of each quantum dot $\epsilon$ (here for simplicity we assume that they are equal) fulfills the resonance condition $\epsilon=\epsilon_{\rm Q}$ with \cite{chirolli2024} 
\begin{equation}\label{Eq:QuartetResonance}
4\epsilon_{\rm Q}+2U+4W=0.
\end{equation}
In such a case, second-order processes involving the exchange of two Cooper pairs with the superconducting leads generate an effective finite quartet coupling $\Gamma_{4e}$ between the vacuum and the fourfold occupied state, as schematically illustrated in Fig.~\ref{fig-outline}(b). However, the repulsive nature of the interactions tends to push these states up in the spectrum, typically at energies higher than those of the doubly occupied states. It follows that, by applying a sufficiently large voltage bias between the normal and superconducting leads, we can probe transitions involving the quartet subspace. In this respect, we study the quartet correlator $Q$ out of equilibrium, and a thorough discussion can be found in Sec.~\ref{Sec:QuartetPhase}.
We now briefly summarize the main results of the present work, which are also briefly outlined in Fig.~\ref{fig-outline}(c)-(f).

\emph{{\it Quartet correlations}}--- In Fig.~\ref{fig-outline}(c), we plot the quartet correlations as a function of dots detuning $\epsilon-\epsilon_{\text{Q}}$. We show that a negligible value of $Q$ at zero bias voltage evolves at finite voltage into a double peak around the resonance condition of Eq.~\eqref{Eq:QuartetResonance} $\epsilon=\epsilon_{\rm Q}$, thus establishing the large contribution of the states Eq.~\eqref{Eq:EquilibriumGS} in the out-of-equilibrium state. Furthermore, even if the phase of the quartet correlator is not a gauge-invariant quantity, as we will discuss in the next sections, the sign change of $Q$ as we sweep the dot level $\epsilon$ across $\epsilon_{\rm Q}$, crossing $Q=0$ exactly at resonance, provides information on the weight of the two states $|\psi^\pm_{\rm Q}\rangle$ of Eq.~\eqref{Eq:EquilibriumGS} in the out-of-equilibrium density matrix. 

A natural question is how to assess the presence of quartet correlations in the system. Focusing on observables that can be directly measured in transport experiments, we study the dissipative Andreev current in the normal terminals, and the current-current correlations quantified by the Fano factor. In addition, we also study the behavior of the dissipationless Josephson current flowing in the superconducting leads for the multi-terminal configuration introduced in Fig.~\ref{fig-outline}(a). The main results are summarized in Fig.~\ref{fig-outline}(d-f). 

\emph{{\it Andreev current.}}--- 
The first observable that we study is the Andreev current in the normal leads. 
The detailed analysis can be found in Sec.~\ref{Sec:Andreev}. We show that, for a voltage bias larger than the quartet splitting, $|\mu|>|\epsilon_{\rm Q}|$, the peak in the Andreev current at the resonant dot level energy $\epsilon_{\rm Q}$ has a width that is given by the quartet gap $|\Gamma_{4e}|$, which is 
the modulus of the quartet coupling $\Gamma_{4e}$. This result shows that we can link the properties of the quartet resonance to a measurable quantity, such as the current-peak linewidth. To further characterize the resonance, we show that, by means of the additional superconducting leads $S_1$ and $S_2$ in Fig.~\ref{fig-outline}(a), we can introduce a dependence of the quartet coupling on the phases $\varphi_1$ and $\varphi_2$, which allows us to modulate the quartet gap in a distinctive way. This ability offers a nonlocal control knob to tune the quartet gap and measure the corresponding change in the width of the Andreev-current peak in an experiment, as shown in Fig.~\ref{fig-outline}(d), providing us with a strong, although indirect, signature of the quartet coupling and enabling a characterization of the quartet gap. 

\emph{{\it Noise signatures.}}--- 
Going one step further, we analyze the current correlations and the Fano factor in the normal terminals, which are discussed in Sec.~\ref{Sec:Fano}. In a hybrid normal-superconducting system, a Fano factor $F=2$ describes a Poissonian emission of Andreev pairs. This is indeed what we find in general away from resonances. Furthermore, close to a {\it pair} resonance, such as those related to local Andreev reflection (LAR) and crossed Andreev reflection (CAR) processes, a rapid exchange of Cooper pairs takes place between the dots and the superconducting leads, which is interrupted by stochastic emission of single electrons, yielding a Fano factor $F=1$ \cite{braggio2011}. This is indeed what we find at the {\it pair} resonances for the Fano factor of the sum of the two dissipative Andreev currents in leads $i=1,2$ for large applied bias voltage, as shown in Fig.~\ref{fig-outline}(e). However, if we decompose the Fano factor $F=1+\Delta F_{1,1}+\Delta F_{1,2}$ down into current-current auto-($1+\Delta F_{1,1}$) and cross-($\Delta F_{1,2}$) correlations contributions, we see that negative cross-correlations are crucial to yield an overall $F=1$ for the sum of the two currents at the LAR resonance, and in the absence of CAR processes (see Sec.~\ref {Sec:Fano}). We explain this phenomenon as an anticorrelation between the currents in the normal leads induced by the common superconducting lead, which coherently exchanges Cooper pairs with one dot at the time. 

This is not the case at the quartet resonance $\epsilon=\epsilon_{\rm Q}$, where, at sufficiently large bias voltage, we find $F=1$ with zero cross-correlation contribution. In this case, two phenomena may occur: the coherent exchange of two Cooper pairs (quartet), or the exchange of a single Cooper pair in the odd-parity sector. The resulting single-electron emission can occur only in a Poissonian manner.  Furthermore, by applying a small detuning of the dot levels from $\epsilon_{\rm Q}$, $\epsilon-\epsilon_{\rm Q}\neq 0$, we find that two peaks appear at the left and right of $\epsilon_{\rm Q}$,  at which the Fano factor reaches values much higher than 2, as shown in Fig.~\ref{fig-outline}(e). This is a strong indication that the steady state of the system has a large contribution from one of the two states in Eq.~\eqref{Eq:EquilibriumGS}. Indeed, a steady state is what survives over long times to all possible single-electron tunneling events (in the sequential tunneling regime). This way, by starting from a steady state dominated by a coherent superposition of states with different numbers of electrons (the vacuum and the fourfold-occupied state), only certain sequences of single-electron tunneling events guarantee a return to the steady state in the long time.  This selectivity creates a trapping phenomenon accompanied by avalanche effects, with a consequent increase in the noise (see Sec.~\ref {Sec:Fano}).   

Furthermore, in the presence of CAR processes, the contributions to the Fano factor from auto-correlations and cross-correlations become equal, apart from a rigid shift of 1 in the former case. This is a rather special case and arises because the CAR processes promote always equal weight superpositions of the dot states and consequently do not distinguish the outgoing normal leads  so that no true distinction can be made between auto- and cross-correlations at very large bias voltage. 

These findings provide evidence for strong many-body correlations in the system, which are linked to finite quartet correlations, and identify the noise in the normal leads as a proxy for the quartet correlations. This signature can be measured in a suitably designed experiment, and its detection would constitute a not trivial {\it smoking gun} signature of quartet correlations in the system.

\emph{{\it Josephson current.}}---
Finally, the setup with three superconducting terminals of Fig.~\ref{fig-outline}(a) allows for the study of the dissipationless Josephson currents flowing in the system. We find that an interesting phenomenology appears at finite bias in the region $\epsilon_{\rm LAR}<\epsilon<\epsilon_{\rm Q}$ between two resonances. Indeed, a well-known tendency of systems driven out of equilibrium in the presence of interactions to induce a sign change of the Josephson current across a given pair resonance \cite{pala2007,governale2008} leads, in the region between two resonances, to a frustration in the system. As a result, as shown in Fig.~\ref{fig-outline}(f), a dominant second harmonic appears in the current-phase relations, which describes a two-Cooper pair current. Unlike what typically arises in $\pi$-periodic junctions obtained in the non-interacting case through interference effects \cite{blatter2001design}, here the derivative of the current at zero phase bias is positive. 

Furthermore, the interaction between the dots gives rise to a non-local Josephson current, which was first pointed out in Ref.~\cite{chirolli2024}, and it is activated at finite bias voltage. 
The non-local Josephson effect describes a Cooper pair drag \cite{peotta2010,peotta2011,semenov2025}, so that when a current bias is applied from terminal $S_0$ to $S_1$, a current is generated in $S_2$ too, despite the latter being kept at the same phase as $S_0$. This phenomenology is partly related to the idea of Cooper quartets as bound pairs of Cooper pairs. At the same time, such a non-local Josephson effect probes only pair correlations, not quartet correlations.

\section{The system} 
\label{Sec:System}

The system we study is composed of two quantum dots, labeled by $i=1,2$, that are tunnel-coupled to two normal leads, $N_1$ and $N_2$, and to one common superconducting lead, $S_0$. In addition, we also consider the possibility of adding two other superconducting leads, $S_1$ and $S_2$, as depicted in Fig.~\ref{fig-outline}(a),  coupled only locally to the nearby quantum dot. If the superconducting phases are all identical $\phi_0=\phi_1=\phi_2$, the setup can be effectively reduced to a configuration with one superconducting terminal and with a renormalized local coupling.  Since we will consider the case in which the coupling with the normal leads is much weaker than the coupling to the superconducting leads, we first study the properties of the unperturbed system, that is with no normal lead. For simplicity, we assume equal tunnel matrix elements $t_{DS}$ between the superconducting leads and the dots.  As far as the common superconducting lead $S_0$ is concerned, the coupling allows Cooper pairs to tunnel to the dots in two different ways: {\it i)} via a local Andreev reflection (LAR) process, where pairs directly tunnel to either one or the other dot, with rate $\gamma_{S}=2\pi|t_{SD}|^2\nu_F$, where $\nu_F$ is the density of states in the lead; or {\it ii)} via a crossed Andreev reflection (CAR) process, where Cooper pairs split up, with one electron going in one dot and the other electron in the other dot, that is controlled by the rate $\gamma_S'=\gamma_S e^{-r_{12}/\xi}\sin(k_Fr_{12})/(k_Fr_{12})$, where $r_{12}=|{\bf r}_1-{\bf r}_2|$ is the distance between the two quantum dots. Note that $|\gamma_S'|\leq |\gamma_S|$ and  $\xi=\hbar v_F/\pi|\Delta|$ is the coherence length in the superconducting lead characterized by a gap $|\Delta|$, $v_F$ and $k_F$ are the carrier Fermi velocity and Fermi momentum, respectively, and $\hbar$ the reduced Planck constant (which we set to one if not explicitly indicated). In turn, the superconducting leads $S_1$ and $S_2$ allow only for LAR processes, controlled by $\gamma_S$, assumed equal for all superconducting leads. We specify the gauge invariant phases $\varphi_i$ of $S_i$ for $i=1,2$ with respect to the reference $S_0$.

In the case $|\Delta|\to \infty$, the operation of integrating away the superconducting leads induces both an effective local pairing on the dots and a non-local singlet pairing between the dots. The effective DQD Hamiltonian then reads \cite{rozhkov2000,meng2009,eldrige2010}, 
\begin{eqnarray}\label{Eq:DDHamiltonian}
    H_{\rm DQD}&=&\sum_{i,\sigma}\epsilon_{i\sigma}n_{i\sigma}+U\sum_in_{i\uparrow}n_{i\downarrow}+Wn_1n_2\nonumber\\
    &-&\sum_{i=1,2}\frac{\gamma_{S,i}}{2}d^\dag_{i\uparrow}d^\dag_{i\downarrow}-\frac{\gamma'_S}{2}(d^\dag_{1\uparrow}d^\dag_{2\downarrow}-{d^\dag_{1\downarrow}d^\dag_{2\uparrow}})+{\rm H.c.}\nonumber\\
\end{eqnarray}with $\gamma_{S,i}=\gamma_S(1+e^{i\varphi_i})$, $n_i=n_{i\uparrow}+n_{i\downarrow}$, $n_{i\sigma}=d^\dag_{i\sigma}d_{i\sigma}$, and  where $d_{i\sigma}$($d^\dag_{i\sigma}$) are dot fermionic annihilation (creation) operators. In the absence of any other perturbation and at equilibrium, all properties of the system can be derived from the ground state of $H_{\rm DQD}$. In the next section, we discuss how quartet correlations are generated for repulsive interactions ($U, W > 0$). For simplicity, we first consider the setup with one superconducting terminal of Fig.~\ref{fig-scheme}(a) and set $\gamma_{S,i}=\gamma_S$.

\subsection{Spectrum and quartet subspace}
\label{Sec:QuartetSubspace}

The Hilbert space of the system is the direct sum of the subspaces with an even and an odd number of electrons (see Appendix \ref{App:basis-states}).  The even parity sector is described by four relevant states, the vacuum $|0\rangle$, the fourfold occupied state $|4e\rangle=d^\dag_{1\uparrow}d^\dag_{1\downarrow}d^\dag_{2\uparrow}d^\dag_{2\downarrow}|0\rangle$, and the two doubly occupied states, the symmetric combination $|d+\rangle=\frac{1}{\sqrt 2}(d^\dag_{1\uparrow}d^\dag_{1\downarrow}+d^\dag_{2\uparrow}d^\dag_{2\downarrow})|0\rangle$ and the spin-singlet state $|S\rangle=\frac{1}{\sqrt 2}(d^\dag_{1\uparrow}d^\dag_{2\downarrow}-d^\dag_{1\downarrow}d^\dag_{2\uparrow})|0\rangle$. 
The other states are not coupled by the effective superconducting pairing terms in Eq.~\eqref{Eq:DDHamiltonian}. 

\begin{figure}[t]
	\centering
	\includegraphics[width=1.0\columnwidth]{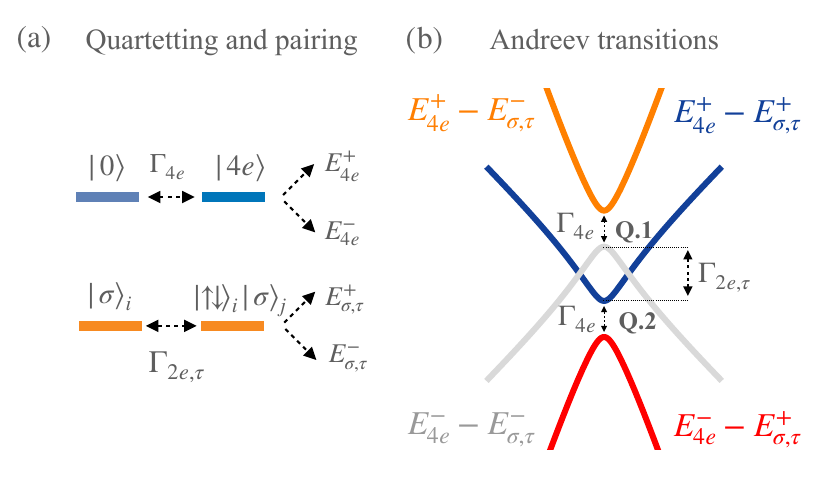}
	\caption{(a) Simple schematics, at the resonance condition $\epsilon=\epsilon_{\rm Q}$, of even- and odd-parity resonance splitting due to $\Gamma_{4e}$ and $\Gamma_{2e}$, respectively. (b)  Andreev transition spectrum at $\epsilon=\epsilon_{\rm Q}$ for the case $\gamma_S'=0$ showing the two avoided crossings \ref{Item-1} and \ref{Item-2}.}
	\label{fig-scheme}
\end{figure}

Close to the resonance condition $\epsilon= \epsilon_{\rm Q}$, the pairing terms generate an admixture of the vacuum and the bare four-electron state with two-electron states, as schematically depicted in Fig.~\ref{fig-scheme}(a). Setting $\epsilon=\epsilon_{\rm Q}$ as a working point, we can construct approximate eigenstates in second-order perturbation theory in the rates $\gamma_S$ and $\gamma_S'$ (see Appendix \ref{App:SWtransformation})
\begin{eqnarray}
\ket{\bar 0}&=&(1-\beta_+)\ket{0}+\beta_-\ket{4e}+\alpha_{\rm lar}|d+\rangle+\alpha_{\rm car}|S\rangle,\label{Eq:0-Eff}~~~\\
\ket{\bar{4e}}&=&(1-\beta_+)\ket{4e}+\beta_-\ket{0}+\alpha_{\rm lar}|d+\rangle-\alpha_{\rm car}|S\rangle,\label{Eq:4e-Eff}~~~~
\end{eqnarray}
where $\alpha_{\rm lar}=\gamma_S/(2\sqrt{2}W)$ and $\alpha_{\rm car}=\gamma_S'/(\sqrt{2}(U+W))$ are first-order coefficients describing the admixture of two-electron states occurring via LAR and CAR  processes, respectively, and $\beta_\pm=(\gamma_S')^2/(4(U+W)^2)\pm \gamma_S^2/(16W^2)$ describe second-order corrections. 

By evaluating the matrix element of the DQD Hamiltonian between the states $|\bar{0}\rangle$ and $|\bar{4e}\rangle$, we obtain the effective quartet coupling $\Gamma_{4e}=\langle\bar{0}|H_{\rm DQD}|\bar{4e}\rangle$, which reads
\begin{equation}\label{Eq:Gamma4e}
\Gamma_{4e}=\frac{\gamma_S^2}{4W}-\frac{(\gamma_S')^2}{2(U+W)}.
\end{equation}
This coupling is the key element that yields eigenstates as coherent superpositions of vacuum and fourfold occupied states, as in Eq.~\eqref{Eq:EquilibriumGS}. We observe that the two contributions from the LAR and CAR processes interfere destructively. The origin of the minus sign is due to the spatial antisymmetry of the Cooper pair wave function in the non-local configuration $|S\rangle$, which in turn originates from the spin-singlet character of the Cooper pairs in the superconducting lead.

By projecting the DQD Hamiltonian onto the quartet subspace, we effectively eliminate first-order terms in $\gamma_S$ and $\gamma_S'$. This procedure can be formalized through a Schrieffer-Wolff transformation (see Appendix~\ref{App:SWtransformation}), and it results in the effective Hamiltonian 
\begin{equation}\label{HeffQuartet}
h^{4e}_{\rm eff}=\Sigma_{4e}+\left(\begin{array}{cc}
0 & \Gamma_{4e}\\
\Gamma^*_{4e} & 4(\epsilon-\epsilon_{\rm Q})
\end{array}\right),
\end{equation}
where $\Sigma_{4e}=\gamma_S^2/(4W)+(\gamma_S')^2/(2(U+W))$ is a self-energy correction.  The effective Hamiltonian for the quartet subspace gives rise to the approximate eigenstates
\begin{equation}
|\pm\rangle_{4e}=\pm u_{4e,\mp}\ket{\bar 0}+e^{-i\theta_{ 4e}}u_{4e,\pm}\ket{\bar{4e}},
\end{equation}
where $\theta_{4e}$ is the phase of the quartet coupling, such that  $\Gamma_{4e}=|\Gamma_{4e}|e^{i\theta_{4e}}$, and $u_{4e,\pm}=\frac{1}{\sqrt{2}}\sqrt{1\pm 2(\epsilon-\epsilon_{\rm Q})/\epsilon_{A,4e}}$, with $\epsilon_{A,4e}=\sqrt{4(\epsilon-\epsilon_{\rm Q})^2+|\Gamma_{4e}|^2}$. The eigenstates have energies $E^\pm_{4e}=2(\epsilon-\epsilon_{\rm Q})+\Sigma_{4e}\pm\epsilon_{A,4e}$. We see that $|\Gamma_{4e}|$ is the effective quartet gap, which splits the degeneracy between the bare vacuum and the fourfold occupied state at the resonance defined by Eq.~\eqref{Eq:QuartetResonance}, as schematically depicted in Fig.~\ref{fig-scheme}(a). 

In turn, exact eigenstates and energy eigenvalues can be written for the odd-parity sector. The latter is decomposed in two, twofold-degenerate, two-dimensional subspaces, spanned by the states $|\sigma\rangle_\tau=\frac{1}{\sqrt 2}(d^\dag_{1\sigma}+\tau d^\dag_{2\sigma})|0\rangle$ and $|\sigma t\rangle_\tau=(|\sigma t\rangle_1+\tau |\sigma t\rangle_2)/\sqrt{2}$, with $\tau=\pm$ and $|\sigma t\rangle_i=d^\dag_{i\sigma}d^\dag_{\bar{i}\uparrow}d^\dag_{\bar{i}\downarrow}|0\rangle$, where $\bar{1}=2$ and $\bar{2}=1$.  Within this subspace, the Hamiltonian reads
\begin{equation}\label{HeffOddParity}
h^o_\tau=\left(\begin{array}{cc}
\epsilon & -\Gamma_{2e,\tau}\\
-\Gamma^*_{2e,\tau} & 3\epsilon-2\epsilon_{\rm Q}
\end{array}\right),
\end{equation}
with $\Gamma_{2e,\tau}=(\gamma_S-\tau\gamma_S')/2$ an effective pairing, whose modulus splits the degeneracy in the odd-parity sector, as schematically depicted in Fig.~\ref{fig-scheme}(a).  The associated eigenstates read 
\begin{equation}\label{OddParity-states}
|\pm\rangle_{\sigma,\tau}=\pm v_{2e,\mp,\tau}\ket{\sigma}_\tau-e^{-i
\theta_{2e,\tau}}v_{2e,\pm,\tau}\ket{\sigma t}_\tau,
\end{equation}
where $\theta_{2e,\tau}$ is the phase of $\Gamma_{2e,\tau}$, such that   $\Gamma_{2e,\tau}=|\Gamma_{2e,\tau}|e^{i\theta_{2e,\tau}}$, and $v_{2e,\pm,\tau}=\frac{1}{\sqrt 2}\sqrt{1\pm (\epsilon-\epsilon_{\rm Q})/\epsilon^o_{A,\tau}}$ 
with $\epsilon^o_{A,\tau}=\sqrt{(\epsilon-\epsilon_{\rm Q})^2+|\Gamma_{2e,\tau}|^2}$. The associated eigenenergies are $E^{o,\pm}_{\sigma,\tau}=2\epsilon-\epsilon_{\rm Q}\pm\epsilon^o_{A,\tau}$. We point out that when $\gamma_S'=0$ the eigenstates are $|\pm\rangle_{\sigma,i}=\pm v_{2e,\mp}\ket{\sigma}_i-e^{-i
\theta_{2e,i}}v_{2e,\pm}\ket{\sigma t}_i$, that are local in the dot index $i=1,2$, whereas for $\gamma_S'\neq 0$ the eigenstates are always symmetric and antisymmetric superpositions of dot states. This difference will become relevant 
when discussing the noise.

\subsection{Setup with three superconducting terminals}
\label{Sec:3SCterminals}

The procedure described in the previous section can be generalized to the setup of Fig.~\ref{fig-outline}(a) featuring three superconducting terminals (see Appendix \ref{App:SWtransformation}). Specifically, in the Hamiltonian Eq.~\eqref{Eq:DDHamiltonian}, the complex pairing amplitude induced by local Andreev reflection reads 
\begin{equation}
\gamma_{S,i}=\gamma_{S}(1+e^{i\varphi_i}), 
\end{equation}
whereas the non-local pairing amplitude $\gamma'_S$ induced by crossed Andreev reflection is left unaltered,  having chosen the superconducting terminal $S_0$ as the reference of phase. As a result, the phases of $S_1$ and $S_2$ enter the quartet coupling  $\Gamma_{4e}=\langle{\bar{0}}|H|\bar{4e}\rangle$, which becomes a complex quantity
\begin{equation}\label{Eq:Gamma4e_phi1phi2}
\Gamma_{4e}=-\frac{(\gamma_S')^2}{2(U+W)}+\frac{\gamma_S^2}{4W}(1+e^{-i\varphi_1})(1+e^{-i\varphi_2}),
\end{equation}
in which we clearly see the different weights of the LAR and CAR contributions and their dependence on the phases $\varphi_i$, thanks to the specific arrangements of leads and their coupling to the quantum dots. The self-energy term is also modified as
\begin{equation}
    \Sigma_{4e}=\frac{\gamma_S^2}{4W}(2+\cos(\varphi_1)+\cos(\varphi_2))+\frac{(\gamma'_S)^2}{2(U+W)},
\end{equation}
and the effective Hamiltonian in the quartet subspace is given by Eq.~\eqref{HeffQuartet} with the proper substitutions.

\subsection{Out-of-equilibrium state} 

In the presence of repulsive interactions, the quartet subspace is pushed up in energy and is not the ground state. To obtain finite dominant quartet correlations, we drive the system out of equilibrium in order to populate also the excited states. To this end, we attach normal leads to the system, which are described by a non-interacting Hamiltonian $H_{\rm leads}=\sum_{i{\bf k}\sigma}\epsilon_{\bf k}c^\dag_{{\bf k}i\sigma}c_{{\bf k}i\sigma}$ and are tunnel-coupled to the quantum dots via the tunneling Hamiltonian \cite{bardeen1962}
\begin{equation}\label{Eq:Htun}
H_{\rm tun}=\sum_{i {\bf k}}t_{ND}c^\dag_{i {\bf k}\sigma}d_{i\sigma}+{\rm H.c.}.
\end{equation}
The full Hamiltonian of the system is given by $H=H_{\rm DQD}+H_{\rm leads}+H_{\rm tun}$. Analogously to the case of the superconducting lead, we define the tunneling rate $\gamma_{N}=2\pi|t_{ND}|^2\nu_F$, which we assume to be the same for both normal leads. We further assume that we are in the transport regime, in which the tunneling rate $\gamma_N$ is the smallest energy scale in the problem, that is, $\gamma_N\ll k_BT\ll \gamma_S,\gamma_S'\ll |\Delta|$. The out-of-equilibrium density matrix in the DQD system can be determined via a master equation formalism \cite{pala2007,governale2008,Nazarov_Blanter_2009}. This procedure is based on a Keldysh perturbative approach \cite{konigPRL1996,konigPRB1996} and a real-time diagrammatic technique \cite{governale2008}, which is a perturbative expansion in the tunneling rate $\gamma_N$. 

In particular, for $\gamma_{N}\ll k_BT$, it is well justified to truncate the expansion in first order, which is also called the sequential tunneling regime, and only consider Fermi golden rule rates $w_{a \leftarrow a'}$. Those quantities describe the transition probability between the eigenstates $|\psi_a\rangle$ of $H_{\rm DQD}$ in Eq.~\eqref{Eq:DDHamiltonian} under the action of the tunneling term Eq.~\eqref{Eq:Htun}. This approach selects the basis of the eigenstates  of $H_{\rm DQD}$ as the relevant one \cite{hussein2016}, so that the density matrix maintains the form $\rho=\sum_aP_a|\psi_a\rangle\langle\psi_a|$, with the populations $P_a$  strongly deviating from the equilibrium ones and determined by the master equation 
\begin{equation}\label{Eq:masterEq}
\dot{P}_a=\sum_{a'}(w_{a\leftarrow a'}P_{a'}-w_{a'\leftarrow a}P_a).
\end{equation}
The steady state solution describing the stationary long-time properties of the density matrix is obtained by setting $\dot{P}_a=0$ and solving for the probabilities $P_a$ with the normalization condition $\sum_a P_a=1$. 

The Fermi golden rule rates account for the possibility that an electron tunnels in or out of the DQD system, 
and they are given by the sum of two terms, which describe the two processes, 
$w_{a \leftarrow a'}=\sum_{i\sigma}w^{i\sigma,{\rm in}}_{a\leftarrow a'}+w^{i\sigma,{\rm out}}_{a\leftarrow a'}$, with
\begin{eqnarray}
w^{i\sigma,{\rm in}}_{a\leftarrow a'}&=&\gamma_{N}f^+(E_a-E_{a'})|\langle\psi_a|d^\dag_{i\sigma}|\psi_{a'}\rangle|^2,\\
w^{i\sigma,{\rm out}}_{a\leftarrow a'}&=&\gamma_{N}f^-(E_{a'}-E_{a})|\langle\psi_a|d_{i\sigma}|\psi_{a'}\rangle|^2,
\end{eqnarray}
where $f^+(\epsilon)=1/(1+\exp((\epsilon-\mu_{N})/k_BT))$ is the Fermi function of the normal leads with chemical potential $\mu_{N}$ (assumed to be equal for all normal leads), $f^-(\epsilon)=1-f^+(\epsilon)$, and $E_a$ is the eigenenergy of the eigenstate $|\psi_a\rangle$. 

Since the tunneling Hamiltonian Eq.~\eqref{Eq:Htun} describes processes in which only single electrons are exchanged between the DQD system and the normal leads, each tunneling event changes the parity of the DQD state. In the long-time limit, a balance is reached between the odd-parity and even-parity sectors.  Importantly, since the Fermi golden rule rates $w_{a\leftarrow a'}$ are first order in $\gamma_N$, the latter factorizes in Eq.~\eqref{Eq:masterEq} and drops out of the steady state probabilities $P_a$'s. This means that the out-of-equilibrium density matrix is of zeroth order in $\gamma_N$ and cannot be obtained perturbatively from the equilibrium one.

\begin{figure*}[t]
	\centering
	\includegraphics[width=2.0\columnwidth]{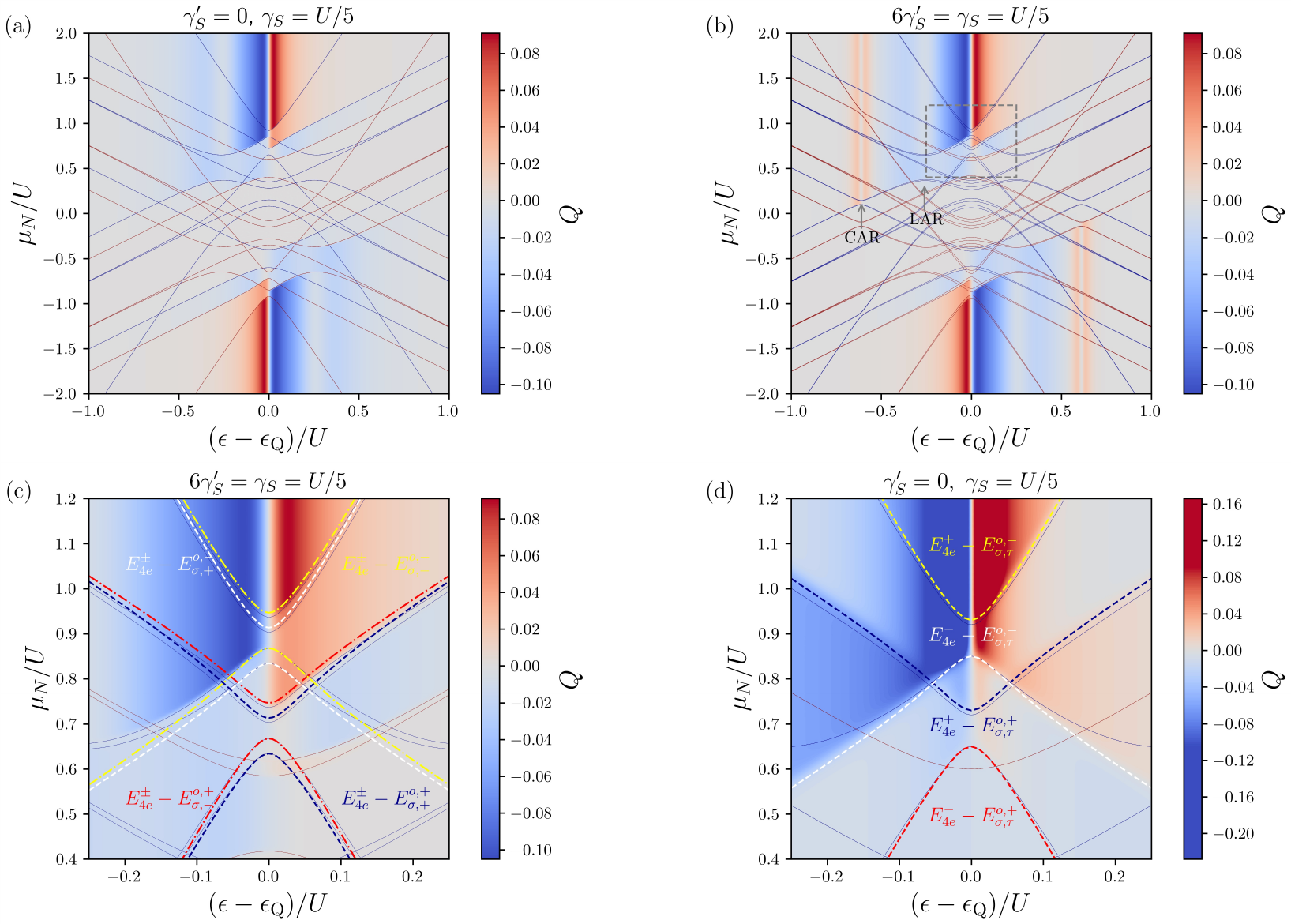}
	\caption{
    (a) Quartet correlator $Q$ calculated on the out-of-equilibrium state as a function of the chemical potential $\mu_N$ of the normal leads and the energy levels of the dots $\epsilon$, for $W=U/4$, $\gamma_S=U/5$ and $\gamma_S'=0$. Shown are also the addition energies $E^e_a-E^o_{a'}$ in blue and $E^o_a-E^e_{a'}$ in red. We consider the case with one common superconducting terminal.  (b) Same as (a) but with $\gamma_S'=\gamma_S/6$. (c) Zoom of the area in (b) shown by the dashed rectangle, where we also show, for comparison, the addition energies derived analytically using the perturbative approach of Sec.~\ref{Sec:QuartetSubspace}, marked as dashed and dot-dashed lines. (d) Same as (c) but with the rates calculated through the effective model of Sec.~\ref{Sec:effModel} for the $\gamma_S'=0$ case. }
	\label{fig-quartet} 
\end{figure*}

\subsection{Andreev transitions spectrum}

The Fermi golden rule rates depend on specific transitions between even- and odd-parity states that give rise to an Andreev transition spectrum, $E_a-E_{a'}$, describing different regions of the transport phase diagram. In the absence of Zeeman terms, setting $\epsilon_{i\sigma}=\epsilon$, and in the absence of interdot tunneling terms, the Andreev transition spectrum and the observables in the out-of-equilibrium case are invariant under the exchange $(\mu_N,\epsilon)\to (-\mu_N,2\epsilon_{\rm Q}-\epsilon)$. The Andreev transition spectrum shows several avoided crossings involving LAR and CAR processes, which have been partly discussed in the literature \cite{hussein2016,hussein2017}. Hereafter, we will focus primarily on the quartet resonances. 

The transitions between the quartet states $|\pm\rangle_{4e}$ and odd-parity eigenstates $|\pm\rangle_{\sigma,\tau}$ describe, for the general case of finite $\gamma_S'$, four avoided crossings, one for each value of $\tau=\pm$ and $\sigma=\uparrow,\downarrow$. Since, in the absence of a Zeeman term, the spectrum is spin-degenerate, each transition is at least doubly degenerate. Additionally, the transitions are degenerate in $\tau$ when $\gamma_S'=0$. This degeneracy is lifted for finite $\gamma_S'$. More precisely, the Andreev-transition spectrum shows the two following main quartet splittings: 
\begin{enumerate}[label=\textbf{Q.\arabic*}]
    \item an avoided crossing $E^\pm_{4e}-E^{o,-}_{\sigma,\tau}$, that describes the energy difference between the quartet states Eq.~\eqref{HeffQuartet}, at energy $E_{4e}^\pm$, and the odd-parity states Eq.~\eqref{OddParity-states}, at energy $E^{o,-}_{\sigma,\tau}=2\epsilon-\epsilon_{\rm Q}-\epsilon^o_{A,\tau}$ and it develops around the chemical potential $\mu_N=(E^+_{4e}+E^-_{4e})/2-E^{o,-}_{\sigma,\tau}= U/2+W+\Sigma_{4e}+\Gamma^\tau_{2e}$; 
    \label{Item-1}
    
    \item an avoided crossing $E^\pm_{4e}-E^{o,+}_{\sigma,\tau}$, describing the energy difference between the quartet states and the odd-parity states at energy $E^{o,+}_{\sigma,\tau}=2\epsilon-\epsilon_{\rm Q}+\epsilon^o_{A,\tau}$, that develops around the chemical potential $\mu_N =(E^+_{4e}+E^-_{4e})/2-E^{o,+}_{\sigma,\tau}= U/2+W+\Sigma_{4e}-\Gamma^\tau_{2e}$. \label{Item-2}
\end{enumerate}
For the case $\gamma_S'=0$, the $\tau$-dependence is lost, and there are two fourfold degenerate avoided crossings, as schematically shown in  Fig.~\ref{fig-scheme}(d). The transitions \ref{Item-1} and  \ref{Item-2} fully characterize the quartet subspace.

\section{Quartet correlations}
\label{Sec:QuartetPhase}

With the out-of-equilibrium steady state density matrix $\rho_{\rm st}$, we can assess several quantities that allow us to fully characterize the system. In particular, we can calculate the quartet correlator $Q$, Eq.~\eqref{Eq:correlatorQ}. The expectation value $\langle\dots\rangle={\rm Tr}[\ldots \rho_{\rm st}]=\sum_aP_a\langle\psi_a|\ldots|\psi_a\rangle$ is taken on the out-of-equilibrium density matrix. The operator $d_{1\downarrow}d_{1\uparrow}d_{2\downarrow}d_{2\uparrow}$ can only have a finite matrix element between states differing by four electrons. The quartet correlator $Q$, whose absolute value is bounded $|Q|\leq 1/2$, is in general complex and its phase reflects the phase of the superposition between $|0\rangle$ and $|4e\rangle$. In the setup featuring a single superconducting terminal, for the gauge choice for which $\gamma_S$ and $\gamma_S'$ are real, $Q$ can take positive or negative values, reflecting which of the two combinations in Eq.~\eqref{Eq:EquilibriumGS} dominates the out-of-equilibrium density matrix, as explained in Sec.~\ref{Sec:Outline}. More generally, any gauge transformation that changes the phase of the fermionic operators results in a phase shift in the quartet correlator. Nevertheless, there are always only two possible eigenstates in the relevant subspace $|0\rangle$, $|4e\rangle$ that have maximal and opposite values of $Q$. In Fig.~\ref{fig-quartet}(a,b), we plot the value of $Q$ as a function of $(\epsilon,\mu_N)$ for $W=U/4$, together with the transition energies $E_a^e-E_a^o$ in blue and $E_a^o-E_a^e$ in red. For simplicity, we consider the setup with a single common superconducting terminal. 

In Fig.~\ref{fig-quartet}(a) we assume $\gamma_S=U/5$ and $\gamma_S'=0$. Close to resonance $\epsilon\simeq \epsilon_{\rm Q}$, we clearly see the onset of finite quartet correlations only when the chemical potential crosses the avoided crossing  \ref{Item-1}. Exactly at $\epsilon=\epsilon_{\rm Q}$, the quartet correlator is zero, and it changes sign when sweeping $\epsilon$ across $\epsilon_{\rm Q}$, as anticipated in Sec.~\ref{Sec:Outline}. Furthermore, in proximity of its node for an energy range of the order $\Gamma_{4e}$ around the resonance, the quartet correlator reaches its maximal value, which, for the chosen parameters, is $Q_{\text{max}}\approx 0.1$. Despite being  20\% of the maximal value of 1/2, which is predicted in the case of attractive interactions \cite{chirolli2024}, it is sufficiently large to influence the behavior of the system. 

In Fig.~\ref{fig-quartet}(b,c), we show the results for $Q$ for finite $\gamma_S'=\gamma_S/6$. The addition energies split due to finite $\gamma_S'$ and four avoided crossings appear, doubly degenerate in the value of $\sigma=\uparrow,\downarrow$. In the close-up view shown in Fig.~\ref{fig-quartet}(c), we show the addition energies obtained from the exact model alongside those calculated using the perturbative approach described in Sec.~\ref{Sec:QuartetSubspace}. The two sets of addition energies exhibit excellent agreement, even for the relatively large value of $\gamma_S/U$ chosen for this figure. Finite quartet correlations appear for values of the chemical potential $\mu_{\rm N}$ larger than the avoided crossing \ref{Item-1}. 

The activation of the quartet correlations when the bias voltage crosses the quartet resonance is a clear indication of the importance of the quartet coupling $\Gamma_{4e}$, which governs the splitting in the quartet sector, as it is clear from inspection of Eq.~\eqref{HeffQuartet}. Maximizing its value makes the superposition Eq.~\eqref{Eq:EquilibriumGS} more robust to small deviations from $\epsilon_{\rm Q}$, thus resulting in stronger quartet correlations. From Eq.~\eqref{Eq:Gamma4e}, we see that the interdot interaction $W$ appears as a fundamental resource to generate quartet correlations, in agreement with what we found in Ref. \cite{chirolli2024}. Furthermore, as we explained in Sec.~\ref{Sec:effModel}, a destructive interference takes place between the contributions originating from LAR and CAR processes. It follows that a way to maximize the quartet gap $\Gamma_{4e}$ is to prevent non-local Andreev processes by setting $\gamma_S'=0$. Nevertheless, the role of a finite $\gamma_S'$ will be discussed in the next sections when considering the current and the current correlations in the normal leads. 

In addition to the main signal which appears at the quartet splitting around $\epsilon_{\rm Q}$, we see that weak quartet correlations appear also close to the LAR resonance around $\epsilon_{\rm LAR}=-U/2$ in Fig.~\ref{fig-quartet}(a,b), and close to the CAR resonance, around $\epsilon_{\rm CAR}=-W/2$ in Fig.~\ref{fig-quartet}(b), and their symmetric partners with opposite 
biases, as given by the general symmetry $(\mu_N,\epsilon)\to (-\mu_N,2\epsilon_{\rm Q}-\epsilon)$.    
These weak quartet correlations are due to the admixture of the two-particle states $|d+\rangle$ and $|S\rangle$ with the vacuum $|0\rangle$ and the fourfold occupied state $|4e\rangle$ in first order at the rates $\gamma_S,\gamma_S'$, analogously to Eqs.~\eqref{Eq:0-Eff} and \eqref{Eq:4e-Eff}. 

\begin{figure*}[t]
	\centering
	\includegraphics[width=2.\columnwidth]{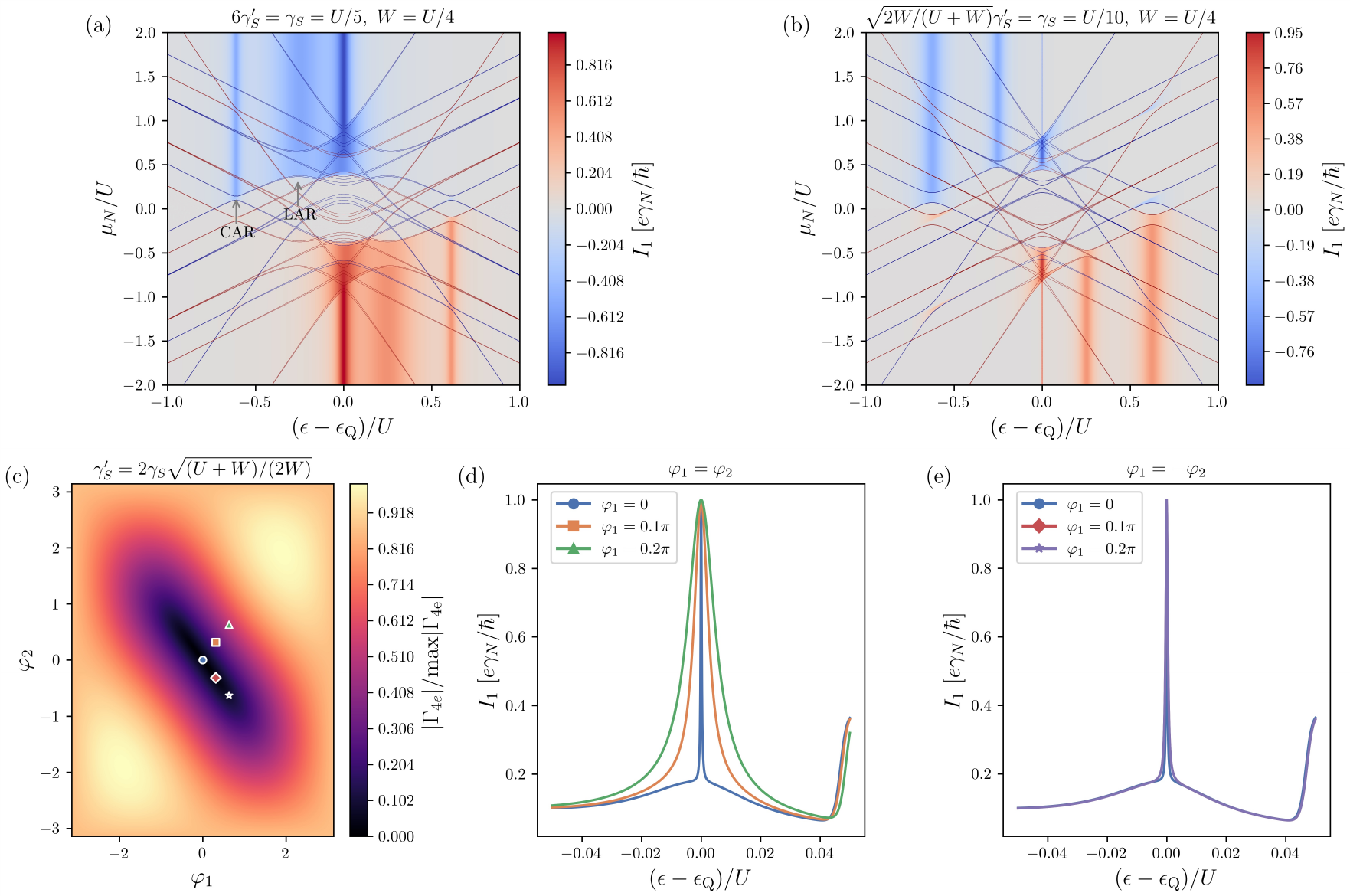}
	\caption{Andreev current $I^A_1$ flowing in normal lead $N_1$. Having chosen an equal DQD configuration, we have that $I_1=I_2$. (a) $I^A_1$ versus chemical potential $\mu_N$ and dot level $\epsilon$ for the case $\gamma_S=U/5$, $W=U/4$, $\gamma_S'=\gamma_S/6$, and  temperature $T=0.0025~U$,  for the  one-superconducting-terminal setup. Shown are also the transition energies between even-parity and odd-parity states and the LAR and CAR resonances. (b)  Current $I_1$ in the  one-superconducting-terminal setup for the case $\gamma_S'=\gamma_S\sqrt{(U+W)/(2W)}$, for which $\Gamma_{4e}=0$. (c) Dependence of the quartet gap $|\Gamma_{4e}|$ Eq.~\eqref{Eq:Gamma4e_phi1phi2} on the phases $\varphi_1$ and $\varphi_2$ in the setup featuring three superconducting terminals, for the condition $\gamma_S'=\gamma_S\sqrt{2(U+W)/W}$ for which $\Gamma_{4e}=0$. (d) Peak in the Andreev current $I^A_1$ at $\mu_N=U$ as a function of $(\epsilon-\epsilon_{\rm Q})/U$ for three different values of the phases along the line $\varphi_2=\varphi_1$ specified in the legend and by the relative symbols, that is refereed to (a). (e) Same as (d) for three different values of the phases along the line $\varphi_2=-\varphi_1$ specified in the legend and by the relative symbols, that is refereed to (c). }
	\label{fig-andreev}
\end{figure*}

\subsection{Reduced model}
\label{Sec:effModel}

For the case $\gamma_S'=0$, we can capture the behavior of the quartet correlator using a simplified model. The odd-parity sector is described by two states, $|\pm\rangle_{\tau,\sigma}$, each one fourfold degenerate in $\tau,\sigma$. Their contributions to the Fermi golden rule rates add incoherently with a prefactor 1/4 accounting for the $\tau$ and $\sigma$ structure. The two states $|\pm\rangle_{\tau,\sigma}$  behave as two effective single levels $|p\rangle_o$, with $p=\pm$. By further considering only the two quartet states $|\pm\rangle_{4e}$ at energy $E^\pm_{4e}$, we can map the problem to the one describing a single-dot system coupled to one superconducting and one normal lead \cite{pala2007,governale2008,braggio2011}.  The resulting Fermi golden rule rates take the expression 
\begin{eqnarray}
w^{oe}_{pp'}&=&\gamma_N|(D^+)^{oe}_{pp'}|^2f^+(E^{oe}_{pp'})+\gamma_N|D^{oe}_{pp'}|^2f^-(-E^{oe}_{pp'}),\nonumber\\
w^{eo}_{pp'}&=&\gamma_N|D^{oe}_{p'p}|^2f^+(E^{oe}_{p'p})+\gamma_N|(D^+)^{oe}_{p'p}|^2f^-(-E^{oe}_{p'p}),\nonumber
\end{eqnarray}
where $E^{oe}_{pp'}=E^o_p-E^{4e}_{p'}$, $p,p'=\pm 1$, and the matrix elements $D^{oe}_{p'p}$ and $(D^+)^{oe}_{p'p}$ are given in App.~\ref{App:RatesEffectiveModel}. The populations ${\bf P}=(P_{4e,+},P_{4e,-},P_{o,+},P_{o,-})$ are obtained by solving the master equation, which, in this four level model, can be done analytically, and the four-operator part of the quartet correlator becomes (see Appendix \ref{App:quartet-correlator})
\begin{equation}\label{Eq:Q-effective-model}
\langle d_{1\downarrow}d_{1\uparrow}d_{2\downarrow}d_{2\uparrow} \rangle=\frac{P_{4e,+}-P_{4e,-}}{2}\frac{\Gamma_{4e}^*}{\sqrt{4(\epsilon-\epsilon_{\rm Q})^2+|\Gamma_{4e}|^2}},
\end{equation}
that has the form of a Breit-Wigner resonance, with the width controlled by the quartet gap $|\Gamma_{4e}|$, and multiplied by the difference of populations. The latter is exactly zero at resonance, reflecting that the vacuum and fourfold-occupied state have equal weight in the eigenstates, and it changes sign on either side of the resonance. The result for the quartet correlator is shown in Fig.~\ref{fig-quartet}(d), and the agreement with the zoomed-in area in Fig.~\ref{fig-quartet}(a) is remarkable, despite the simplification. The four-point expectation value depends on the quartet coupling in its complex form, as expected.

\section{Andreev currents} 
\label{Sec:Andreev}
 
The system schematized in Fig.~\ref{fig-outline}(a) clearly supports Andreev currents flowing in the normal leads. We can then express the current operator through lead $j$ as $\hat{I}_j=-e d\hat{N}_j/dt=-i(e/\hbar)[\hat{H}_{{\rm tun},j},\hat{N}_j]$. Due to the general form of the tunneling Hamiltonian and the DQD Hamiltonian, the expectation value of the current is expressed as
\begin{equation}
I_j=\frac{ie}{\hbar}t_{ND}\sum_{{\bf k}\sigma}\langle c^\dag_{j{\bf k}\sigma}d_{j\sigma}\rangle+{\rm c.c.},
\end{equation} 
and it is taken as a trace over the out-of-equilibrium density matrix. The current, and also the higher current cumulants such as the noise, can be calculated elegantly by introducing counting fields $\chi_j$ in the Fermi golden rule rates, $w(\boldsymbol{\chi})_{a \leftarrow a'}=\sum_{j\sigma}w^{j\sigma,{\rm in}}_{a\leftarrow a'}e^{-i\chi_j}+w^{j\sigma,{\rm out}}_{a\leftarrow a'}e^{i\chi_j}$, within the full counting statistics framework. Following this procedure, the current in the normal leads can be written as \cite{braggio2006,flindt2008,flindt2010,braggio2011} (see also Appendix \ref{App:Noise}),
\begin{equation}
I^A_j=-i\frac{e}{\hbar}\sum_{a,a'}\left.\frac{\partial w_{a \leftarrow a'}(\boldsymbol\chi)}{\partial\chi_j}\right|_{\boldsymbol\chi=0}P_{a'},
\end{equation}
where $P_{a}$ are the populations in the out-of-equilibrium stationary steady state.

The Andreev current through the system has been extensively studied, both from a theoretical \cite{ishizaka1995,choi2000,martin-rodero2011,braggio2011,hussein2016} and an experimental point of view \cite{deacon2015,amitai2016,tan2021,kurtossy2022,debbarma2023,dejong2023}, with particular reference to Cooper pair splitting \cite{recher2001,recher2003,sauret2004quantum,chevallier2011current,burset2011microscopic,hussein2016,brange2021dynamic,brange2024adiabatic,delfino2025cooper-pair}. There are several avoided crossings in the Andreev transition spectrum, that give rise to resonances in the Andreev current at $\epsilon_{\rm LAR}$, $\epsilon_{\rm CAR}$, and $\epsilon_{\rm Q}$, and that have been previously partly characterized in the literature \cite{hussein2016}. Here, we will concentrate on the signatures of the quartet correlations discussed in Fig.~\ref{fig-quartet}. 

In Fig.~\ref{fig-andreev} we study the Andreev current $I_1$ flowing in lead 1, that is equal to the current flowing in lead 2, $I_1=I_2$, having chosen, for simplicity, a symmetric configuration between leads and dots. With reference to Fig.~\ref{fig-andreev}(a), at $\epsilon=\epsilon_{\rm Q}$ there are two main contributions: {\it i)} a signal arising from LAR processes that becomes resonant around $\mu_N=\pm \epsilon_{\rm LAR}$ \footnote{This resonant behavior originates from the pair splitting in the odd-parity sector, as detailed in Sec.~\ref{Sec:QuartetSubspace}.}, and {\it ii)} a second signal that is activated at higher chemical potential, when crossing the quartet-dominated avoided crossings \ref{Item-1} and \ref{Item-2}.

Focusing on this second signal, we see that the current first increases in magnitude around the quartet splitting \ref{Item-2}, then it persists at the same magnitude for larger chemical potential  $\mu_N>E^+_{4e}-E^-_{\sigma,-}$ above \ref{Item-1}, but the width of the peak as a function of the level position $\epsilon$ shrinks.  The width of the narrow peak around $\epsilon_{\rm Q}$ is of the order of the quartet gap $|\Gamma_{4e}|$. As we previously mentioned, the LAR and CAR processes that contribute to the quartet coupling interfere destructively. Therefore, we expect the linewidth to disappear when $\gamma_S'$ reaches the specific value $\gamma_S\sqrt{(U+W)/(2W)}$ which yields  $\Gamma_{4e}=0$ in Eq.~\eqref{Eq:Gamma4e}. This destructive-interference effect is shown in Fig.~\ref{fig-andreev}(b), where we notice that the linewidth shrinks substantially, although it does not completely vanish, due to higher-order processes that do not fully cancel via destructive interference. This result strongly links the resonance in the Andreev current at large bias with finite quartet correlations, suggesting an experimentally viable way to directly measure the strength of quartet correlation, i.e., the gap $|\Gamma_{4e}|$.  We notice that this result is obtained in the setup with only one superconducting terminal and, since $|\gamma_S'|\leq|\gamma_S|$, it requires the condition  $|W|>|U|$, which is typically not realized. In the next section, we will discuss the setup with three superconducting terminals, which allows us to tune the quartet gap by means of the phases of the different superconducting leads.

\subsection{Tuning the Andreev peak}

As explained in Sec.~\ref{Sec:QuartetSubspace}, the expression for $\Gamma_{4e}$ shows that the contributions from $\gamma_S$ and $\gamma_S'$ tend to cancel due to the destructive interference between the LAR and CAR processes that contribute to the quartet splitting. 

An appealing way to tune the quartet coupling $\Gamma_{4e}$ is to use the phases of two additional superconducting leads, $S_1$ and $S_2$, in the setup of Fig.~\ref{fig-outline}(a). The phases $\varphi_1$ and $\varphi_2$ hence act as control knobs for $\Gamma_{4e}$ (see  Eq.~\ref{Eq:Gamma4e_phi1phi2}). This mechanism was also proposed in Ref.~\cite{chirolli2024} (see Appendix \ref{App:two-terminal} for the simple two-terminal case).

The simplest possibility is to study the dependence of the high-bias Andreev peak at $\epsilon_{\rm Q}$ for the usual case $|\gamma_S'|< |\gamma_S|$ and by setting $\varphi_1=\varphi_2$. The last condition is equivalent to a setup with two superconducting terminals, as discussed in App.~\ref{App:two-terminal}. The result is shown in Fig.~\ref{fig-outline}(d), where the width of the peak is shown as a function of the detuning $\epsilon-\epsilon_{\rm Q}$ for a few values of the phase $\varphi_1=\varphi_2$. The width of the peak clearly varies between a maximum and a minimum, and studying this dependence may allow the characterization of the quartet gap. At the same time, we notice that the peak at the LAR resonance also varies similarly. 

To emphasize the characteristics of the quartet gap we focus on the three-terminal setup. In Fig.~\ref{fig-andreev}(c) we plot the quartet gap $|\Gamma_{4e}(\varphi_1,\varphi_2)|$, normalized to its maximum value, as a function of the two phase differences for the particular case $\gamma_S'=\sqrt{2(U+W)/W}\gamma_S$, for which the quartet gap vanishes at  $\varphi_1=\varphi_2=0$. Although this condition is hardly realized in experiments, we consider it as the best-case scenario to tune the quartet gap. We see that along the two orthogonal directions $\varphi_1=\pm \varphi_2$, the quartet gap varies in a different way. In Fig.~\ref{fig-andreev}(d) we plot the peak of the Andreev current around the resonance $\epsilon_{\rm Q}$ and for $\mu_N=U$, along the line $\varphi_1=\varphi_2$ and in Fig.~\ref{fig-andreev}(e) along the line $\varphi_1=-\varphi_2$. We see that the width of the peak reflects the variation of the gap in the $(\varphi_1,\varphi_2)$ plane shown in Fig.~\ref{fig-andreev}(c), and as such it is a tool to experimentally characterize the quartet correlations.

Despite the dependence on the phases $\varphi_1$ and $\varphi_2$, the destructive interference between the contributions originating from the LAR and CAR processes persists in the expression of the quartet coupling Eq.~\eqref{Eq:Gamma4e_phi1phi2}. We point out that a strategy to modify 
the term originating from CAR processes could be to employ Rashba spin-orbit\cite{hussein2017} coupled superconducting leads or magnetically active barriers between the dots and the superconducting leads, which would add an induced pairing in the spin-triplet channel in the DQD Hamiltonian, making the physical mechanisms even richer.

\begin{figure*}[t]
	\centering
	\includegraphics[width=2.0\columnwidth]{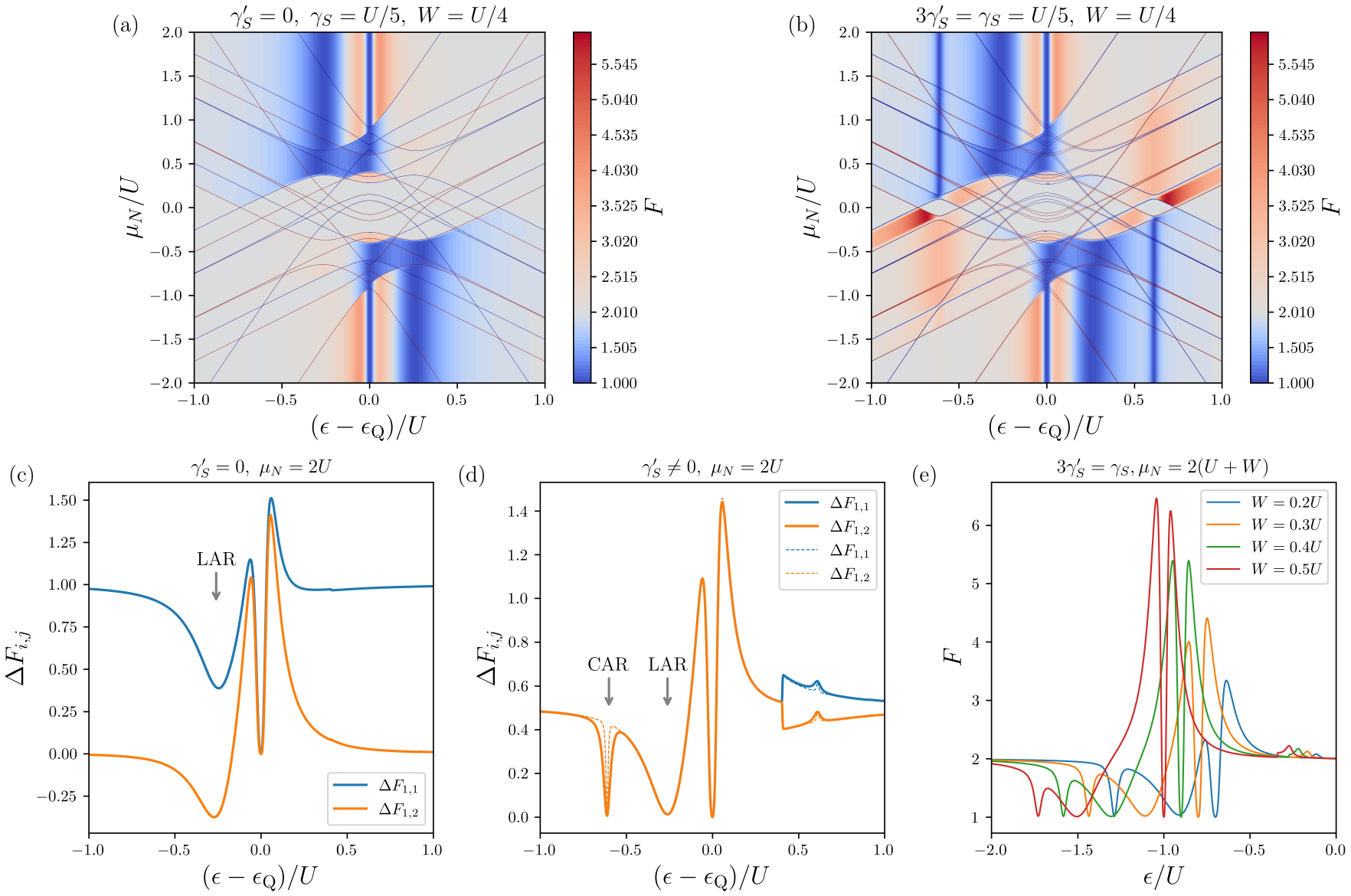}
	\caption{Fano factor $F=\sum_{ij}S_{ij}/|e\sum_jI_j|$ accounting for the auto- and cross-correlation noise in the two normal leads for the setup of Fig.~\ref{fig-outline}(a) for $\varphi_1=\varphi_2=0$ and for (a) $\gamma'_S=0$ and (b) $\gamma_S'=\gamma_S/3$. (c,d) Breakdown of the Fano factor in contributions accounting for auto- and cross-correlations, $\Delta F_{i,j}=S_{ij}/|e I_i|-\delta_{i,j}$, evaluated for (c) $\gamma_S'=0$ and (d) $\gamma'_S=\gamma_S/3$ (continuous line) and $\gamma'_S=0.1\gamma_S$ (thin dashed lines).  Parameters for (a-d) are $W=U/4$, $\gamma_S=U/5$. (e) Fano factor $F$ at $\mu_N=2(U+W)$ as a function of $\epsilon$ for several values of $W/U$, while keeping $\gamma_S/U$ and $\gamma_S'/U$ fixed. }
	\label{fig-fano}
\end{figure*}

\section{Fano factor}
\label{Sec:Fano}

Besides the average current sustained by Andreev reflection processes, it is also interesting to consider the noise in the normal leads, in particular the Fano factor. The symmetrized zero-frequency noise is the current-current correlator 
\begin{equation}
S_{ij}=\int dt \langle\{\delta\hat{ I}_i(t),\delta\hat{I}_j(0)\}_+\rangle,
\end{equation}
with $\delta\hat{ I}_i=\hat{I}_i-\langle\hat{I}_i\rangle$ the current fluctuation in the $i$-th terminal and $\{,\}_+$ the anticommutator. This correlator can be calculated through second derivatives of the cumulant generating function, or via a recursive scheme outlined in Refs.~ \cite{braggio2006,flindt2008,flindt2010,braggio2011,walldorf2020noise} (see Appendix \ref{App:Noise}). The Fano factor is the noise divided by the product of the current and the electron charge $e$. We consider the two normal-lead configuration of Fig.~\ref{fig-outline}(a). Since the tunneling processes which determine the noise fluctuations in each normal metal lead are independent in the lowest order in the tunneling, we start studying the Fano factor of the sum of the currents $I_1$ and $I_2$, which reads
\begin{equation}
F=\frac{\sum_{i,j=1,2}S_{ij}}{e\left|\sum_jI_j\right|}.
\end{equation}

In general, in superconducting systems and away from resonances, the Fano factor takes the value $F=2$, describing Poissonian emission of Andreev pairs \cite{muzykantskii1994}. However, in systems with interactions, the Fano factor can take different values, signaling strong correlations \cite{braggio2003,novotny2004,kock2005}. 
Additionally, as pointed out in \cite{braggio2011}, at resonance between the vacuum and the twofold occupied states, the superconducting proximity effect leads to fast coherent oscillations of a Cooper pair between the dots and the superconductor, that are stochastically interrupted by injection of single electrons from the normal lead via uncorrelated single-electron tunneling events, leading to the universal Poissonian Fano factor $F=1$. 

The results for the Fano factor are presented in Fig.~\ref{fig-fano}, where we consider for simplicity the setup of Fig.~\ref{fig-outline}(a) without the additional superconducting lead $S_1$ and $S_2$. In Fig.~\ref{fig-fano}(a) the Fano factor is plotted as a function of $\epsilon$ and $\mu_N$ for $\gamma_S'=0$. We see that in the Coulomb blockade regime and away from resonances the Fano factor takes the value $F=2$ (gray background), indicating that electrons are transferred as Andreev pairs.

Furthermore, we see that exactly at the resonances $\epsilon_{\rm LAR}$ and $\epsilon_{\rm Q}$ we have $F=1$ (blue). At $\epsilon_{\rm LAR}$ we interpret the result as originating from fast local Cooper pair exchange between the DQD system and the superconducting lead, as in Ref.~\cite{braggio2011}. 
In turn, in the region around $\epsilon_{\rm Q}$, we see a qualitative change of the Fano factor across the quartet splitting \ref{Item-1}, where it reaches values much larger than  2, for bias voltage larger than the quartet resonance, $\mu_N>(E^+_{4e}+E^-_{4e})/2-E^{o,-}_{\sigma,\tau}= U/2+W+\Sigma_{4e}+\Gamma^\tau_{2e}$. By a direct comparison of Fig.~\ref{fig-fano}(a) with Fig.~\ref{fig-quartet}(a) we see a very similar behavior between the quartet correlator and the Fano factor. This, in itself, suggests that the Fano factor is a possible proxy for the quartet correlations. 

In Fig.~\ref{fig-fano}(b) we consider finite $\gamma_S'=\gamma_S/6$. Far away from resonances and in the Coulomb blockade diamond, the Fano factor is $F=2$, as expected. Exactly at the resonances $\epsilon_{\rm Q}$, $\epsilon_{\rm LAR}$, and $\epsilon_{\rm CAR}$ the Fano factor is $F=1$, where, in addition to the cases previously discussed, the finite $\gamma_S'$ opens a further transport channel via crossed Andreev reflection. In the regions where the quartet correlator is finite, the Fano factor reaches values similar to the case $\gamma_S=0$ of Fig.~\ref{fig-fano}(a). It is important to point out, that when we consider a zero (finite) $\gamma_S'$, what we actually mean is $\gamma_S'$ much smaller (greater) than $\gamma_N$, the tunneling rate to the normal leads. This means that the results obtained within the sequential tunneling approximation at finite $\gamma_S'$ will not smoothly connect with the results obtained for $\gamma_S'=0$, as the intermediate regime when $\gamma_S'$ is of the same order of $\gamma_N$ is not described by the sequential tunneling approximation.

It is interesting to study separately the contributions to the Fano factor coming from the current-current auto- and cross-correlations in the normal leads. For the symmetric setup considered here, we have $S_{1,1}=S_{2,2}$ and $S_{1,2}=S_{2,1}$, together with $\langle \hat{I}_1\rangle=\langle \hat{I}_2\rangle$. We then study $\Delta F_{i,j}\equiv S_{i,j}/|e I_i|-\delta_{i,j}$, where we subtract 1 from the auto-correlation Fano factor to single out the contributions of the different processes, so that we write $F=1+\Delta F_{1,1}+\Delta F_{1,2}$.

\subsection{Absence of CAR processes, $\gamma_S'=0$}

\begin{figure}[t]
	\centering
	\includegraphics[width=1.0\columnwidth]{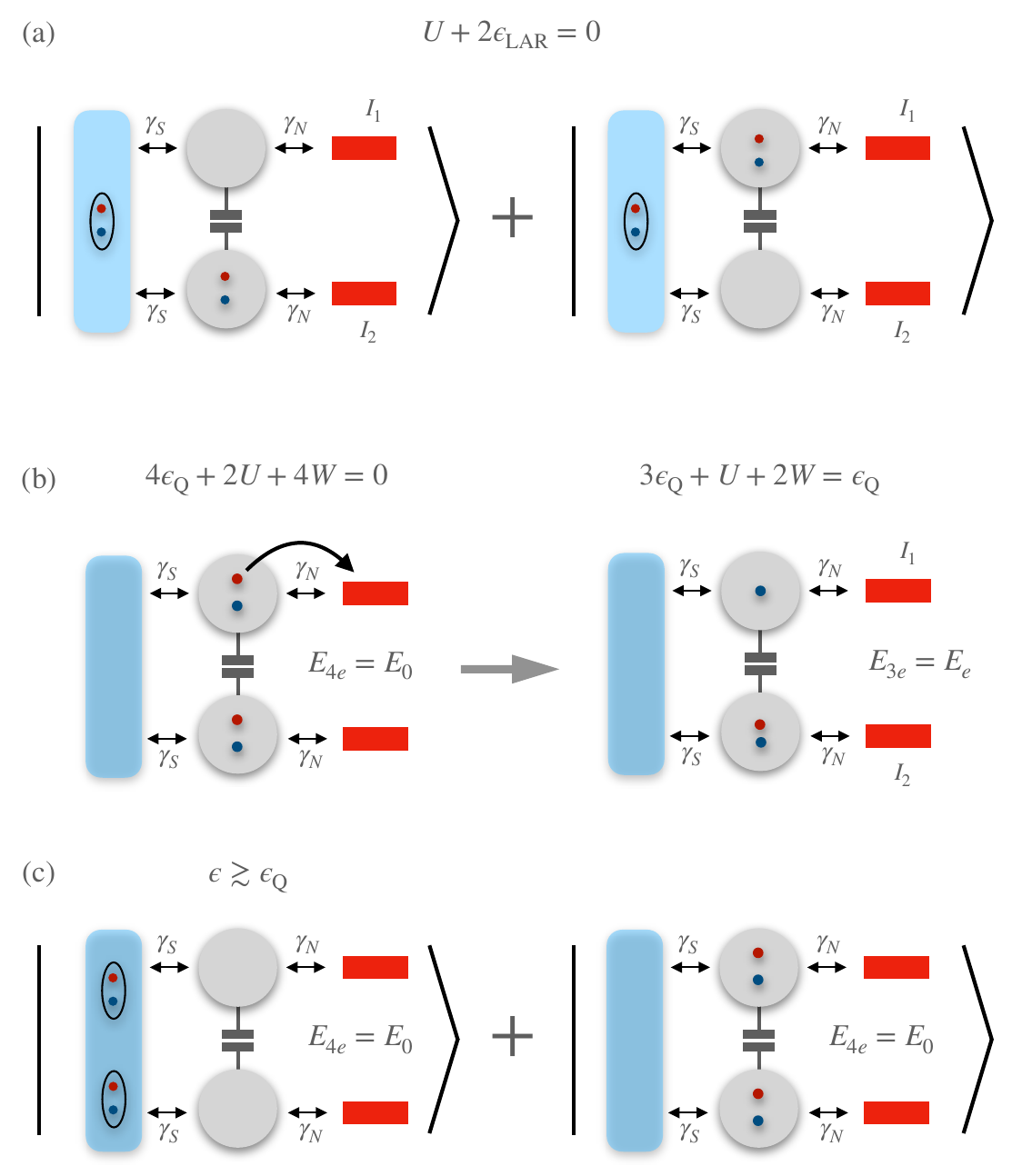}
	\caption{(a) Schematics of the state $|d+\rangle$ at $\epsilon_{\rm LAR}$ as a coherent superposition of a Cooper pair in one dot or in the other. The possible single-electron tunneling events necessarily involve only the lead coupled to the dot containing the Cooper pair, leading to a void in the short-time-scale current in the other lead, and a consequent negative cross-correlation. (b) Steady state at $\epsilon_{\rm Q}$ as a purely mixed states, leading to uncorrelated single-electron tunneling events and $F=1$. (c) Steady state at weak detuning $\epsilon\gtrsim \epsilon_{\rm Q}$ as a coherent superposition yielding finite quartet correlations. The tunneling of a single electron is followed by the fast ejection of three electrons, so as to recover the steady state in the long time, leading to a super-Poissonian Fano factor, with equal auto- and cross-correlation contributions.}
	\label{fig-noise}
\end{figure}

The result for the case in which $\gamma_S'=0$ is shown in Fig.~\ref{fig-fano}(c) for high bias. Away from the resonances, cross-correlations are negligible. At $\epsilon_{\rm LAR}$, we notice that the total Fano factor $F=1$ arises from the addition of negative cross-correlations to auto-correlations exceeding unity, signaling that the stochastic Poissonian single electron emission in the sum of the two currents $I_1$ and $I_2$ is the result of a more complex anticorrelated behavior. This process can be understood as the result of Cooper pair hopping to the DQD from the same superconducting lead in the presence of a repulsive interdot interaction $W$. Indeed, exactly at $\epsilon_{\rm LAR}$, the vacuum state $|0\rangle$ is resonantly coupled, by the local effective pairing in DQD Hamiltonian, to the symmetrized state $|d+\rangle=\frac{1}{\sqrt{2}}(d^\dag_{1\uparrow}d^\dag_{1\downarrow}+d^\dag_{2\uparrow}d^\dag_{2\downarrow})|0\rangle$, that is a coherent superposition of a Cooper pair in one or the other dot, as schematized in Fig.~\ref{fig-noise}(a). Fast Cooper pair exchange between the superconducting lead and the DQD generates, in the short time-scale, coherence between the vacuum and $|d+\rangle$, selecting one of the two possible states $|\pm\rangle_{2e}=\frac{1}{\sqrt 2}(|0\rangle\pm|d+\rangle)$. Exactly at resonance $\epsilon_{\rm LAR}$, the stationary state resulting from coupling to the normal leads in the sequential tunneling regime has equal probability of the two degenerate states $|\pm\rangle_{2e}$, thus resulting in  $\rho_{\rm st}\simeq \frac{1}{2}(|0\rangle\langle 0|+|d+\rangle\langle d+|)$. To understand the noise, we need to consider all possible sequences of single-electron tunneling processes for which the system starts from and returns to the steady state. Every single-electron tunneling taking place in lead $j=1,2$ projects the system into the state with one electron in dot $j$. Consequently, the next single-electron tunneling event can occur only from the dot $j$, leading to a current hiatus in the other lead. As a result, the single-electron tunneling events are not completely random but become anticorrelated between the two leads, resulting in an effective antibunching phenomenon characterized by negative cross-correlations. 

In turn, exactly at $\epsilon_{\rm Q}$ in Fig.~\ref{fig-fano}(c),  we see the total absence of cross-correlations and a perfect matching between the auto-correlation and the cross-correlation contributions in a region on order $\sim|\Gamma_{4e}|$ around $\epsilon_{\rm Q}$. In particular, we observe positive cross-correlations away from $\epsilon_{\rm Q}$. Exactly at $\epsilon_{\rm Q}$, the out-of-equilibrium steady state has equal populations of the states Eq.~\eqref{Eq:EquilibriumGS}, and of the odd-parity states Eq.~\eqref{OddParity-states},  
\begin{equation}
\rho_{\rm st}=P_{4e}(|0\rangle\langle 0|+|4e\rangle\langle 4e|)+P_{\sigma}(|\sigma\rangle\langle \sigma|+|\sigma t\rangle\langle \sigma t|),
\end{equation}
where we dropped the $\tau$-label for simplicity and with $P_{4e}$ and $P_\sigma$ we denote the population of the relevant even- and odd-parity sectors, respectively. The net result is a purely mixed state, schematized in Fig.~\ref{fig-fano}(b), characterized by zero quartet correlations, as explained in Sec.~\ref{Sec:QuartetPhase}. The purely mixed character of the steady state promotes uncorrelated single-electron tunneling events, with zero auto- and cross-correlations and Fano factor $F=1$. 

By applying a small detuning $\delta \epsilon =\epsilon-\epsilon_{\rm Q}$, the steady state changes to
\begin{equation}
\rho_{\rm st}\approx\begin{cases}
    P^+_{4e}|+\rangle_{4e}{}_{4e}\langle +|+P^+_\sigma |+\rangle_{\sigma}{}_\sigma\langle +|, \qquad \delta \epsilon\sim \Gamma_{4e},\\
P^-_{4e}|-\rangle_{4e}{}_{4e}\langle -|+P^-_\sigma |-\rangle_{\sigma}{}_\sigma\langle -|,\qquad  \delta \epsilon\sim -\Gamma_{4e},
\end{cases}
\end{equation} 
as schematized in Fig.~\ref{fig-fano}(c). Now, similarly to the case discussed at $\epsilon_{\rm LAR}$, the coherent superposition forces some sequences of single-electron tunneling events and forbids others, because the net initial and final state is always the steady state. In particular, the coherent quartet coupling $\Gamma_{4e}$ mediates the fast exchange of two Cooper pairs with the superconducting lead. Once an electron tunnels out of one of the dots, the system is projected onto the state with three electrons, which does not appear in the steady state, so that an avalanche effect takes place in the short time-scale, with three electrons ejected to the leads, leading to strongly correlated transport characterized by positive cross-correlations $\Delta F_{1,2}$ equal to the contribution $\Delta F_{1,1}$ to the auto-correlations and super Poissonian noise.  We interpret the large Fano factor in close proximity of $\epsilon_{\rm Q}$ as a strong manifestation of the correlated behavior induced by the interactions and a {\it smoking gun} signature of quartet correlations, thus identifying the noise as a proxy for the quartet correlator, as previously pointed out. These are clear, experimentally accessible features that can witness the presence of strong quartet correlations in the system.

At low bias, where quartet correlations are negligible, we report that the $F=1$ result is characterized by negligible cross-correlations (not shown), both at $\epsilon_{\rm LAR}$ and $\epsilon_{\rm Q}$, with the latter arising from the pair resonance between the singly occupied and threefold occupied states in the odd-parity sector.

\subsection{Presence of both LAR and CAR processes}

A unique result appears when studying the contributions of auto- and cross-correlations separately for the case of finite $\gamma_S'$. In Fig.~\ref{fig-fano}(d), we see that, for large positive bias and on the left of the resonances $E^+_{4e}-E^{o,-}_{\sigma,\tau}$, the auto-correlations and the cross-correlations are exactly equal, apart from the rigid shift of 1 in the lead Fano factor. In particular, we notice that a finite, albeit small, value of $\gamma_S'$ (much larger than $\gamma_N$) is sufficient to render the curves equal. This result follows from the fact that the presence of finite $\gamma_S'$ puts on the same level the two mechanisms of LAR and CAR for Cooper pair exchange between the dots and the superconducting lead, by promoting equal weight superposition of the two dots as eigenstates in the odd-parity sector, $|\sigma\rangle_\tau$ and $|\sigma t\rangle_\tau$. It then follows that no true distinction can be made between single-electron processes taking place in one or the other dot/lead.  

This result can be understood also by noticing that at sufficiently large and positive bias and on the left of the resonances $E^+_{4e}-E^{o,-}_{\sigma,\tau}$ all the Fermi functions are equal to one,  $f^+(E_a-E_{a'})=1$, so that we have $w(\boldsymbol{\chi})_{a \leftarrow a'}=\sum_{j\sigma}w^{j\sigma,{\rm in}}_{a\leftarrow a'}e^{-i\chi_j}$. It then follows that the general matrix of rates $W(\boldsymbol{\chi})$ satisfy a Poissonian structure, i.e. $(\partial^n W(\boldsymbol{\chi})/\partial (-i\chi_j)^n)|_{\boldsymbol{\chi}=0}=W^{\rm in}_j$ (see Appendix \ref{App:Noise}). Following Ref.~\cite{flindt2010}, for the symmetric DQD configuration we are considering, with the same bias $\mu_N$ applied to leads 1 and 2, and for the case $\gamma_S'\neq 0$, the two matrices $W_1^{\rm in}=W_2^{\rm in}$ are equal, so that $\Delta F_{1,1}=\Delta F_{1,2}$ (see Appendix \ref{App:rates}). 

We also study the Fano factor around the $\epsilon_{\rm Q}$ resonance as a function of the interdot interaction strength $W$, while keeping $U$ and all the other parameters constant. As shown in Fig.~\ref{fig-fano}(e), the Fano factor maintains a similar form, with minima where it goes down to $F=1$ at $\epsilon_{\rm Q}$, $\epsilon_{\rm LAR}$, and $\epsilon_{\rm CAR}$, and at the same time the maxima around $\epsilon_{\rm Q}$ reach high values which increase with increasing $W$. We interpret this phenomenology as the signature of avalanche dynamics triggered by quartet correlations that determine super-Poissonian noise features \cite{flindt2004,koch2005}. We point out that, even if the increase of the Fano factor to a value up to $F\sim 4$ could be suggestive of charge-$4e$ carriers, we do not think it is correct to argue in such a way. Indeed, we also report Fano factors that exceed 4, reaching non-universal, non-integer values. 

Finally, we point out that a strong noise signal characterized by large values of the Fano factor appears also at very low biases $\mu\approx\gamma_S'$ for $\epsilon\gtrsim \epsilon_{\rm CAR}$, as shown in Fig.~\ref{fig-fano}(b). This signal is not associated to the quartet physics since in that region of the transport diagram we have negligible quartet correlations, $Q\approx 0$. We ascribe this super Poissonian noise to the blockade and lifting mechanism associated with the singlet state $|S\rangle$, similar to the Coulomb blockade lifting  \cite{nguye2006}. Indeed, starting from a state with one electron per dot at the CAR resonance, fast Cooper-pair exchange can take place between the dots and the SC lead. After the tunneling of one electron, the DQD remains trapped in an odd-parity state that can be lifted only by a tunneling from the other dot. This leads to rapid bursts and long silences, resulting in super-Poissonian noise.

\begin{figure*}[t]
	\centering
	\includegraphics[width=2.0\columnwidth]{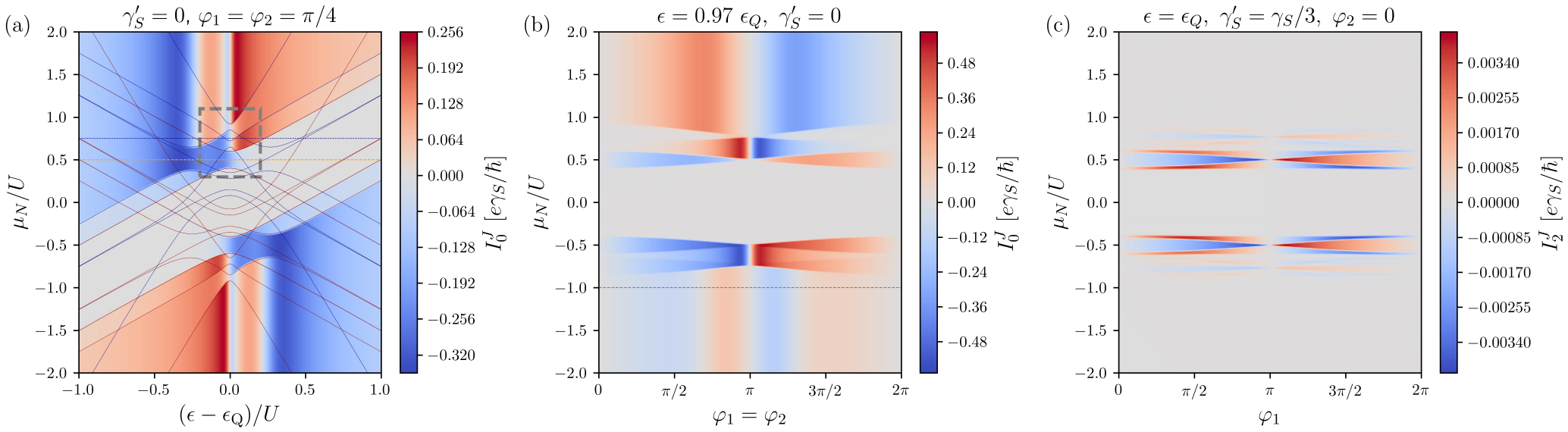}
	\caption{ Josephson currents $I^J_i$ for the setup of Fig.~\ref{fig-outline}(a). (a) $I^J_0$ in lead $S_0$ as a function of $\epsilon/U$ and $\mu_N/U$, for phase differences $\varphi_1=\varphi_2=\pi/4$ and for $\gamma_S'=0$.  (b) $I^J_0$ as a function of the phase difference $\varphi_1=\varphi_2$ and $\mu_N$, for $\epsilon=0.97\epsilon_{\rm Q}$  and $\gamma_S'=0$. (c) Non-local Josephson current $I^J_2$ in lead $S_2$ as a function of $\varphi_1$, for $\gamma_S'=\gamma_S/3$, $\epsilon=\epsilon_{\rm Q}$. In all panels $\gamma_S=U/5$. The two dashed lines in (a) and the one in (b) indicate the voltages at which the current-phase relations in Fig.~\ref{fig-outline}(f) are obtained.}
	\label{fig-josephson}
\end{figure*}

\section{Josephson current}

We conclude the analysis of the system by studying the dissipationless Josephson currents flowing into the system in the three-superconducting-terminal setup of Fig.~\ref{fig-outline}(a). In the sequential tunneling regime that we are considering, the out-of-equilibrium density matrix is at zeroth order in the tunneling rate $\gamma_N$, and the resulting Josephson current is also at zeroth order in $\gamma_N$. This means that the Josephson (dissipationless) current at the lowest (zeroth) order in $\gamma_N$ is conserved between the superconducting terminals, such as $\sum_i I^J_i=0$, and we can limit the sum only on the index of superconducting lead $S_i$ with $i=0,1,2$. 

We fix the reference phase $\phi_0=0$ and assign the gauge invariant phase-differences $\varphi_i$ to the superconducting terminals $S_i$, which are in one-to-one correspondence with the dots $i$, with $i=1,2$.  The Josephson currents are then given by 
\begin{equation}\label{Eq:I_Jeta}
I^J_i=-\frac{2e\gamma_S}{\hbar}{\rm Im}[e^{i\varphi_i}\langle d_{i\downarrow}d_{i\uparrow}\rangle],
\end{equation}
and $I^J_0=-(I_1^J+I^J_2)$.  This well-known expression shows that the Josephson current is related to the {\it pair} correlation function, and we do not expect it to provide direct information on the quartet correlations. In particular, the expectation value of pair operators on the states Eq.~\eqref{Eq:EquilibriumGS} is zero. Nevertheless, we will discuss two effects that are related to the combination of interactions and non-equilibrium effects, namely the ability to support two-Cooper pair currents and a non-local Josephson effect.

\emph{{\it Current versus $\mu_N$ and $\epsilon$.}}--- 
As a first step, we characterize the Josephson current $I^J_0$ in the common superconducting lead $S_0$. This is shown in Fig.~\ref{fig-josephson}(a) as a function of $(\epsilon-\epsilon_{\rm Q})/U$ and $\mu_N/U$, for $\varphi_1=\varphi_2=\pi/4$ and for the case $\gamma'_{S}=0$. In the conventional case and from the choice of phases of Fig.~\ref{fig-outline}(a), we expect that a positive Josephson current flows from $S_0$ to $S_j$ for positive values of the phase $\varphi_j$. This way, when $I^J_0>0$ we have a "$\pi$-junction" effect. In this respect, we note that since the system is expected to exhibit high-harmonic content in the current-phase relation, we will use the terminology "$\pi$-junction" in a loose way, simply to refer to the fact that a positive phase bias yields a negative current around zero phase difference. From Fig.~\ref{fig-josephson}(a), we see an overall $\pi$-junction behavior when sweeping $\epsilon$ across the resonances, which is of the same kind as those described in Refs.~\cite{samuelsson2000,pala2007} and that is due to the out-of-equilibrium condition in the presence of interactions. In particular, a Josephson current becomes finite when the bias $\mu_N$ crosses a pair splitting in the Andreev transition spectrum. In the region where quartet correlations are relevant, we only see an increase in the signal. In addition, the particular configuration of superconducting terminals, for which CAR processes are active only between the dots and $S_0$, does not yield a signal at the CAR resonance (not shown), because Cooper pairs coming from $S_0$ that split in the two dots cannot be transferred to the two other superconducting electrodes $S_i$, so they do not contribute to the Josephson current.

\emph{{\it Two-Cooper pair current.}}---
Nevertheless, the presence of two interacting quantum dots between the superconducting terminals generates a current-phase relation with a high harmonic content. Looking at Fig.~\ref{fig-josephson}(a) in the region between $\epsilon_{\rm LAR}$ and $\epsilon_{\rm Q}$ the system is frustrated between two opposite $\pi$-junction tendencies. As a result, the current acquires a dominant second harmonic. A second harmonic in the current-phase relation is  a signature of  two-Cooper pair Josephson current and in this case arises as an out-of-equilibrium effect.  This is appreciated in Fig.~\ref{fig-josephson}(b), where the current crosses $I^J_0=0$ for three values of the phase $\varphi_1=\varphi_2$ as we move slightly away from $\epsilon_{\rm Q}$ in the region between the two resonances, that is for negative bias $\mu_N$ [see the dashed lines in Fig.~\ref{fig-josephson}(a) that are also plotted in Fig.~\ref{fig-josephson}(b)].   This is also seen in Fig.~\ref{fig-outline}(f), where we show the current $I^J_1+I^J_2$ as a function of $\varphi_1=\varphi_2$ for three cases: for $\epsilon=\epsilon_{\rm Q}$, $\mu_N=-\epsilon_{\rm CAR}$, for $\epsilon=\epsilon_{\rm Q}$, $\mu_N=-\epsilon_{\rm Q}$, and for $\epsilon=0.97~\epsilon_{\rm Q}$, $\mu_N=-U$. Furthermore, the dominant second harmonic has a positive sign, a condition that is rarely encountered. Indeed, the $\pi$-periodic Josephson current-phase relations that are obtained in the rhombus configuration \cite{blatter2001design,doucot2002pairing,ioffe2002topologically,ioffe2002possible,doucot2003topological,doucot2005protected,kitaev2006protected,doucot2012physical,gladchenko2009superconducting,bell2014protected,bondar2025} or in hybrid superconducting devices \cite{larsen2020parity-protected,messelot2024phase,arnault2025multiplet,leblanc2025gate,banszerus2024voltage-controlled,ciaccia2024charge-4e,banszerus2025hybrid} are based on an interference effect, in which the first harmonic, $\sin(\varphi)$, cancels. The second harmonic has opposite sign, $-\sin(2\varphi)$, so that the current flows in the direction opposite to the phase bias, around zero phase-difference.

\emph{{\it Non-local Josephson effect.}}---
In Ref.~\cite{chirolli2024}, a non-local Josephson effect was pointed out, which is activated at the quartet resonance in a setup with attractive interactions. The key observation is that the presence of the CAR-induced term in the quartet coupling $\Gamma_{4e}$ in Eq.~\eqref{Eq:Gamma4e_phi1phi2} with no phase dependence yields a non-trivial dependence of the energy and the currents on the phases. In particular, we can obtain a finite current $I^J_2(\varphi_1)$ as a function of $\varphi_1$ for $\varphi_2=0$. In other words, a phase bias between the terminals $S_0$ and $S_1$ also yields a current in $S_2$. This effect corresponds to a Cooper pair drag \cite{peotta2010,peotta2011,semenov2025} and is activated only at finite $\gamma_S'$. The effect appears for bias voltages $\mu_N$ and dot level $\epsilon$ in the dashed rectangle in Fig.~\ref{fig-josephson}(a), but for the case of finite $\gamma_S'$. 
To understand this, we notice that it is a Josephson current, so it becomes active when the bias $\mu_N$  crosses a pair resonance. 
Furthermore, in the region marked by the dashed rectangle, the finite $\gamma_S'$ yields further splitting of the Andreev addition energies, and for intermediate values of the normal lead chemical potentials, the Fermi functions do not have definite values, with many channels opening and closing as we change the bias voltage.

\section{Summary and Discussion}

In this work, we studied the possibility of inducing superconducting quartet correlations in a system composed of two quantum dots with repulsive interactions, coupled to superconducting and normal leads, and driven out of equilibrium by a bias voltage. As pointed out in Ref.~\cite{chirolli2024}, a coherent coupling of strength $\Gamma_{4e}$ between the vacuum state and the fourfold occupied state is generated in second order in the induced pairing on the dots, when the resonance condition Eq.~\eqref{Eq:QuartetResonance} is met. The latter gives rise to a finite quartet correlator $\langle d_{1\downarrow}d_{1\uparrow}d_{2\downarrow}d_{2\uparrow}\rangle-\langle d_{1\downarrow}d_{1\uparrow}\rangle\langle d_{2\downarrow}d_{2\uparrow}\rangle-\langle d_{1\downarrow}d_{2\uparrow}\rangle\langle d_{1\uparrow}d_{2\downarrow}\rangle$ that is zero in absence of interactions. For attractive interactions, as in the case discussed in Ref.~\cite{chirolli2024}, the quartet correlator acquires a maximum value in the ground state. However, for repulsive interactions, the quartet manifold is pushed up in energy in the spectrum, and the ground state is characterized by weak quartet correlations. We suggest bringing the double-quantum-dot system out of equilibrium through a bias voltage applied to normal leads, tunnel-coupled to the quantum dots. In this way, an out-of-equilibrium population of the quartet subspace is achieved, which we show leads to finite quartet correlations. 

We discuss several observables, such as the Andreev current, the current-current correlations, and the associated Fano factor in the normal leads. In particular, we show that a high-bias peak at the quartet resonance $\epsilon\approx\epsilon_{\rm Q}$ in the Andreev current has a width that is given by the quartet gap $|\Gamma_{4e}|$, that can be tuned by the phase of additional superconducting leads. In particular, we show that we can modify the interference between the LAR and CAR processes through appropriate phase modulation. The resulting complex, nontrivial behavior yields a unique transport footprint of the quartet phenomenology. Those phase-dependent properties may also be useful in specific applications, such as parametric amplifiers or even phase-controlled electronics relevant to the field of quantum technologies.

When studying the noise, we find that the total Fano factor through the normal leads closely mirrors the activation of the quartet correlator at high voltage bias, and it can reach values $F>2$ in close proximity to the quartet correlator maxima. Furthermore, when focusing on the current-current auto- and cross-correlation contributions to the Fano factor in the two normal leads, we observe a unique behavior at high bias voltage, where the two contributions are identical, apart from a unitary shift, in a window of detuning of width $|\Gamma_{4e}|$ around the quartet resonance. We find this behavior a possible {\it smoking gun} of the presence of strong quartet correlations in the system. 

Finally, we show that the device also exhibits an intriguing phenomenology in the Josephson currents in a configuration with three superconducting leads, characterized by a current-phase relation dominated by the second harmonic in a specific region of the transport phase diagram, where quartet correlations play a role, and that it can support a non-local Josephson effect characterized by Cooper-pair drag. All these features may also be attractive for potential applications in the growing field of quantum technologies.

In conclusion, the phenomenology uncovered opens novel and unprecedented avenues for the study of higher-order superconductivity in hybrid superconducting devices driven out of equilibrium. More generally, this work outlines a direction for engineering exotic states of matter in systems composed of conventional elements. Additional interesting phenomenology can emerge by including also  spin-orbit interaction, Zeeman coupling, or triplet superconductivity in the leads. Further developments are necessary to probe the system more precisely and obtain more direct signatures of quartet correlations. Possible extensions may include the study of finite-frequency noise, waiting-time distributions, time-dependent phenomenology, Shapiro-step signatures, and photon-assisted quartet correlations induced in hybrid superconducting setups coupled to microwave radiation.

\section{Acknowledgments}
The authors acknowledge useful discussion with R. Fazio. L.C. acknowledges funding from Fondazione Cariplo under the grant N. 2023-2594.
A.B. acknowledges the 
MUR-PRIN 2022—Grant No. 2022B9P8LN-(PE3)-Project NEThEQS “Non-equilibrium coherent thermal effects in quantum systems” in PNRR Mission 4-Component 2-Investment 1.1 “Fondo per il Programma Nazionale di Ricerca e Progetti di Rilevante Interesse Nazionale (PRIN)” funded by the European Union-Next Generation E, the project "Thermoelectric effects in solid-state quantum devices based on multiterminal Josephson junctions” of the bilateral agreement CNR/CONICET (Italy/Argentina) 2026-2027 and CNR project QTHERMONANO.

\appendix

\section{Basis states}
\label{App:basis-states}

In Table \ref{tab.:HS_basis} we provide the basis states with the convention used in the text, by means of which we can write the general eigenstates $|\psi_a\rangle$ as
\begin{eqnarray}\label{EqApp:States-general}
    |\psi^e_a\rangle&=&C_{a,0}|0\rangle+\sum_{i=1,2}C_{a,di}|\!\uparrow\downarrow\rangle_i+C_{a,S}|S\rangle+C_{a,4e}|4e\rangle\nonumber\\
    &+&C_{a,T0}|T0\rangle+\sum_{\sigma=\uparrow,\downarrow}C_{a,T\sigma}|T\sigma\rangle,\\
    |\psi^o_a\rangle&=&\sum_{i,\sigma}C_{a,i\sigma}|\sigma\rangle_i+C_{a,i\sigma t}|\sigma t\rangle_i,
\end{eqnarray}

\begin{table}[ht]
		\begin{ruledtabular}
			\begin{tabular}{lll}
				$\ket{0}$ & & vacuum \\
				$\ket{S}$ & $=\frac{1}{\sqrt{2}} \big( 
				d^\dag_{1\uparrow} d^\dag_{2\downarrow} 
				-d^\dag_{1\downarrow} d^\dag_{2\uparrow}
				\big)\ket{0}$ & singlet state\\
				$\ket{\uparrow\downarrow}_i$ & $= d_{i\uparrow}^\dag  d_{i\downarrow}^\dag \ket{0}$ & doubly occupied states\\
				$\ket{4e}$ & $=d_{1\uparrow}^\dag  d_{1\downarrow}^\dag d_{2\uparrow}^\dag  d_{2\downarrow}^\dag \ket{0}$
				& fourfold occupied state\\
				$\ket{T0}$ & $=\frac{1}{\sqrt{2}}\big( 
				d^\dag_{1\uparrow} d^\dag_{2\downarrow}
				+d^\dag_{1\downarrow} d^\dag_{2\uparrow}
				\big)\ket{0}$ & unpolarized triplet state\\
				$\ket{T\sigma}$ & $=d^\dag_{1\sigma} d^\dag_{2\sigma}\ket{0}$ & polarized triplet states\\
				\hline
				$\ket{\sigma}_i$ & $=d^\dag_{i\sigma}\ket{0}$ & singly occupied states\\
				$\ket{\sigma t}_i$ & $=d^\dag_{i\sigma}d^\dag_{\bar{i}\uparrow}d^\dag_{\bar{i}\downarrow}\ket{0}$
				& triply occupied states
			\end{tabular}
		\end{ruledtabular}
		\caption{\label{tab.:HS_basis} Basis states used in the text, divided into even parity (top rows) 
		and odd parity (bottom rows). }
	\end{table}

\section{Schrieffer-Wolff transformation}
\label{App:SWtransformation}

Here, we provide the derivation of the effective quartet Hamiltonian. We separate the DQD Hamiltonian as $H_{\rm DQD}=H_0+V$, with an unperturbed term describing number-conserving terms (quantum dot energies), $H_0$, and a perturbation $V$, that describes the coupling with superconducting leads with pairing terms proportional to $\gamma_S$ and $\gamma_S'$. We can eliminate $V$ at first order in $\gamma_S$ and $\gamma_S'$ through a unitary transformation $e^SH_{\rm DQD}e^{-S}$ implemented by the unitary operator $e^{-S}$, where $S$ is anti-Hermitian and of the same order of $V$. Expanding the transformed Hamiltonian at second order in the perturbation or the operator $S$ we have
\begin{eqnarray}
e^S(H_0+V)e^{-S}&=&H_0+V+[S,H_0]+[S,V]\nonumber\\
&+&\frac{1}{2}[S,[S,H_0]]+\ldots.
\end{eqnarray}
By choosing $S$ to satisfy $[S,H_0]+V=0$ we are left with the effective Hamiltonian
\begin{equation}
H_{\rm eff}=H_0+\frac{1}{2}[S,V].
\end{equation}
Such a procedure is particularly useful when a given low-energy subspace is separated from other higher energy states by a large energy scale. In the present case, this scale is approximately given by ${\rm min}(U,W)$. The last step is to project the Hamiltonian on the desired, low-energy subspace. 

This procedure is also known as quasi-degenerate perturbation theory. We can introduce many-body eigenstates of the unperturbed term, such that $H_0=\sum_n\epsilon^0_n|n\rangle\langle n|+\sum_l\epsilon^0_l|l\rangle\langle l|$, with $\{|n\rangle\}$ and $\{|l\rangle\}$ spanning the low-energy and high-energy subspaces, respectively, and $V=\sum_{nl}|n\rangle\langle l|V_{nl}+{\rm H.c.}$, with $V_{nl}=\langle n|V|l\rangle$. It then follows that 
\begin{equation}
S=-\sum_{nl}|n\rangle\langle l|\frac{V_{nl}}{\epsilon^0_l-\epsilon^0_n}-\sum_{nl}|l\rangle\langle n|\frac{V_{ln}}{\epsilon^0_n-\epsilon^0_l},
\end{equation}
and it diagonalizes the Hamiltonian at first order in the perturbation. Including the second order correction to the unperturbed term, the matrix elements of the effective Hamiltonian, projected onto the subspace spanned by $\{|n\rangle\}$, read
\begin{equation}\label{App:Heff}
\langle n'|H_{\rm eff}|n\rangle=\epsilon^0_n\delta_{n,n'}+\sum_lV_{n'l}V_{ln}\frac{\epsilon^0_l-(\epsilon^0_n+\epsilon^0_{n'})/2}{(\epsilon^0_n-\epsilon^0_l)(\epsilon^0_l-\epsilon^0_{n'})}.
\end{equation}
The same result can be obtained by standard perturbation theory. Indeed, we identify the eigestates at second order as
\begin{eqnarray}
|n\rangle^{(2)}&=&(1-S+\frac{1}{2}S^2)|n\rangle\nonumber\\
&=&|n\rangle+\sum_l|l\rangle\frac{V_{ln}}{\epsilon^0_n-\epsilon^0_l}+\sum_{n'l}|n'\rangle\frac{V_{n'l}V_{ln}/2}{(\epsilon^0_n-\epsilon^0_l)(\epsilon^0_l-\epsilon^0_{n'})},\nonumber\\
\end{eqnarray}
so that by projecting the Hamiltonian on the eigenstates at second order, ${}^{(2)}\langle n'|(H_0+V)|n\rangle^{(2)}$ and keeping only contributions up to second order in $V$  we obtain Eq.~\eqref{App:Heff}.

\subsection{Effective Hamiltonian at $\epsilon=\epsilon_{\rm Q}$}
\label{App:EffectiveH-at-resonance}

Our goal is to obtain an effective Hamiltonian for the quartet subspace. Here, we provide the procedure for the most general case with the three-superconducting terminal configuration and to this end we relabel as $\phi_i$ the phase of the superconducting lead $S_i$, with $i=0,1,2$. This way, in the Hamiltonian Eq.~\eqref{Eq:DDHamiltonian}, the LAR-induced pairings $\gamma_{S,i}$ acquire a dependence on the phases $\varphi_i$,
\begin{equation}
    \gamma_{S,i}=\gamma_{S_0,d_i}e^{i\phi_0}+\gamma_{S_i,d_i}e^{i\phi_i},
\end{equation}
where $\gamma_{S_i,d_j}$ is the bare rate for tunneling between superconducting lead $S_i$ and dot $j$. Similarly, the CAR-induced pairing acquires the form
\begin{equation}
    \gamma_S'=\gamma'_{S_0,d_i} e^{i\phi_0}, 
\end{equation}
with $\gamma'_{S_0,d_i}=\gamma_{S_0,d_i}e^{-r_{12}/\xi}\sin(k_Fr_{12})/(k_Fr_{12})$. The relevant states are  $\{\ket{0},\ket{4e},|S\rangle,|\!\uparrow\downarrow\rangle_1,|\!\uparrow\downarrow\rangle_2\}$, with the convention that $|\!\uparrow\downarrow\rangle_i=d^\dag_{i\uparrow}d^\dag_{i\downarrow}|0\rangle$, and the Hamiltonian in this subspace reads
\begin{equation}\label{Eq:H4x4}
h_e=\left(\begin{array}{ccccc}
0 & 0 &  -\frac{(\gamma'_S)^*}{\sqrt{2}} & -\frac{\gamma^*_{S1}}{2} & -\frac{\gamma^*_{S2}}{2} \\
0 & 4(\epsilon-\epsilon_{\rm Q}) & \frac{\gamma_S'}{\sqrt{2}} & -\frac{\gamma_{S2}}{2} & -\frac{\gamma_{S1}}{2}\\
-\frac{\gamma'_S}{\sqrt{2}} & \frac{(\gamma'_S)^*}{\sqrt{2}} & 2\epsilon+W & 0 & 0\\
-\frac{\gamma_{S1}}{2} & -\frac{\gamma^*_{S2}}{2} & 0 &  2\epsilon+U & 0\\
-\frac{\gamma_{S2}}{2} & -\frac{\gamma^*_{S1}}{2} & 0 & 0 &  2\epsilon+U
\end{array}\right).
\end{equation}
For $\epsilon=\epsilon_{\rm Q}=-U/2-W$, the two high-energy states $|0\rangle$ and $|4e\rangle$ are separated by the low-energy states $|\!\uparrow\downarrow\rangle_i$ and $|S\rangle$ by an energy scale on order of $U,W$ and the perturbation $V$ connects them. The eigenstates at second order in $\gamma_{Si}$ and $\gamma_S'$ are 
\begin{eqnarray}
\ket{\bar{0}}&=&\left[1-\beta_+\right]\ket{0}+\beta_-\ket{4e}\nonumber\\
&-&\frac{\gamma_{S1}}{4W}|\!\uparrow\downarrow\rangle_1-\frac{\gamma_{S2}}{4W}|\!\uparrow\downarrow\rangle_2-\frac{\gamma_S'}{\sqrt{2}(U+W)}|S\rangle,\label{Eq:App:bar0}\\
\ket{\bar{4e}}&=&\left[1-\beta_+\right]\ket{4e}+\beta^*_-|0\rangle\nonumber\\
&-&\frac{\gamma^*_{S2}}{4W}|\!\uparrow\downarrow\rangle_1-\frac{\gamma^*_{S1}}{4W}|\!\uparrow\downarrow\rangle_2-\frac{\gamma_S'^*}{\sqrt{2}(U+W)}|S\rangle,\label{Eq:App:bar4e}
\end{eqnarray}
where
\begin{eqnarray}
\beta_+&=&\frac{|\gamma_{S1}|^2+|\gamma_{S2}|^2}{32W^2}+\frac{|\gamma_S'|^ 2}{4(U+W)^2}\nonumber\\
\beta_-&=&\frac{1}{4}\left[\frac{\gamma_S'^2}{(U+W)^2}-\frac{\gamma_{S1}\gamma_{S2}}{4W^2}\right].
\end{eqnarray} 
The effective Hamiltonian in the basis $\{\ket{\bar{0}},\ket{\bar{4e}}\}$ reads
\begin{equation}
h^{4e}_{\rm eff}=\left(\begin{array}{cc}
\Sigma_{4e} & \Gamma_{4e}\\
\Gamma^*_{4e} & 4(\epsilon-\epsilon_{\rm Q})+\Sigma_{4e}
\end{array}\right),
\end{equation}
where
\begin{eqnarray}
\Gamma_{4e}&=&\frac{\gamma^*_{S1}\gamma^*_{S2}}{4W}-\frac{(\gamma'^*_S)^2}{2(U+W)},\\
\Sigma_{4e}&=&\frac{1}{8W}(|\gamma_{S1}|^2+|\gamma_{S2}|^2)+\frac{|\gamma'_S|^2}{2(U+W)}.
\end{eqnarray}
By setting $\phi_0=0$, $\phi_1=\varphi_1$, and $\phi_2=\varphi_2$ we have $\gamma_{S,i}=\gamma_S(1+e^{i\varphi_i})$ and real $\gamma_S'$ and we recover the results presented in Sec.~\ref{Sec:3SCterminals}.

\section{Two-terminal configuration}
\label{App:two-terminal}

The simplest configuration featuring two superconducting terminal is a specular configuration, with a second terminal $S_0'$ coupled to the right of the quantum dots in Fig.~\ref{fig-outline}(a), in exactly a specular way. Assuming for simplicity inversion symmetry of the entire setup, the modification amounts to the  substitution in the Hamiltonian Eq.~\eqref{Eq:DDHamiltonian} of the rates as
\begin{equation}
    \gamma_S\to 2\gamma_S \cos(\varphi/2),\qquad
    \gamma'_S\to 2\gamma'_S \cos(\varphi/2),
\end{equation}
with $\varphi$ the phase difference between the  superconducting leads on the right and on the left. Consequently, all quantities associated to the quartet subspace acquire a straightforward dependence on the phase difference. In particular, the quartet coupling reads
\begin{equation}
\Gamma_{4e}(\varphi)=4\cos^2(\varphi/2)\left[\frac{\gamma_S^2}{4W}-\frac{(\gamma_S')^2}{2(U+W)}\right].
\end{equation}
The dependence of $\Gamma_{4e}$ on the external phase difference $\varphi$ allows us to tune the width of the peak in the Andreev current and thus obtain information about the quartet gap.

\section{Effective rates in the reduced model}
\label{App:RatesEffectiveModel}

Here, we provide the effective rates of the reduced $4\times 4$ model described in Sec.~\ref{Sec:effModel}. The golden rule matrix elements that connect the even sector states $\ket{\pm}_{4e}$ with the fourfold odd states $\ket{\pm}_{\tau,\sigma}$ with $\tau=\pm$ and $\sigma=\uparrow,\downarrow$ can be computed analytically. We have
\begin{eqnarray}
{}_{\tau\sigma}\langle p'|d_{is}|p\rangle_{4e}&=&\frac{1}{2}(\delta_{i1}+\tau \delta_{i2})(\delta_{s\uparrow}\delta_{\sigma\downarrow}-\delta_{s\downarrow}\delta_{\sigma\uparrow})D^{o\tau,e}_{p',p},\nonumber\\
{}_{\tau\sigma}\langle p'|d^\dag_{is}|p\rangle_{4e}&=&\frac{1}{2}(\delta_{i1}+\tau \delta_{i2})\delta_{s\sigma}(D^+)^{o\tau,e}_{p',p},
\end{eqnarray}
with $p,p'=\pm$ and 
\begin{eqnarray}
D^{o\tau,e}_{p',p}&=&p'B_\tau(-p') pA(-p)(\alpha_{\rm lar}+\tau\alpha_{\rm car}) \nonumber\\
&+& p'B_\tau(-p')  A(p)(\alpha_{\rm lar}-\tau\alpha_{\rm car})\nonumber\\
    &-& \sqrt{2} B_\tau(p') [pA(-p)\beta_-+A(p)(1-\beta_+)]\\
(D^+)^{o\tau,e}_{p',p}&=&-B_\tau(p') p A(-p)(\alpha_{\rm lar}-\tau\alpha_{\rm car})\nonumber\\
&-& B_\tau(p') A(p)(\alpha_{\rm lar}+\tau\alpha_{\rm car}))  \nonumber\\
    &+& p'\sqrt{2} B_\tau(-p') (pA(-p)(1-\beta_+)+A(p)\beta_-)\nonumber\\
\end{eqnarray}
with $A(\pm)=u_{4e,\pm}$ and $B_\tau(\pm)=v_{2e,\pm}$. For $\gamma_S'=0$, we have that $\alpha_{\rm car}=0$ and the dependence on $\tau$ is lost in the expressions of the matrix elements relevant for the Fermi golden rule rates.

\subsection{Quartet correlator in the reduced model}
\label{App:quartet-correlator}

Here, we provide details of the calculation of the quartet correlator in the framework of the reduced model. We make use of the expressions Eqs.~\eqref{Eq:App:bar0},\eqref{Eq:App:bar4e} for the states $|\bar{0}\rangle$ and $|\bar{4e}\rangle$, and the expressions for the eigenstates in the quartet sector $|\pm\rangle_{4e}=\pm u_{4e,\mp}|\bar{0}\rangle+e^{-i\theta_{4e}}u_{4e,\pm}|\bar{4e}\rangle$. We then find
\begin{eqnarray}
    {}_{4e}\langle\pm|d_{1\downarrow}d_{1\uparrow}d_{2\downarrow}d_{2\uparrow}|\pm\rangle_{4e}&=&\pm \frac{(\Gamma_{4e}\beta_-^2+\Gamma_{4e}^*(1-\beta_+)^2)}{2\sqrt{4(\epsilon-\epsilon_{\rm Q})^2+|\Gamma_{4e}|^2}}\nonumber\\
    &+&\beta_-(1-\beta_+).
\end{eqnarray}
By noticing that $|\Gamma_{4e}|\sim \beta_-\sim \beta_+$, at lowest order in $\gamma_S,\gamma_S'$, we obtain
\begin{equation}
    {}_{4e}\langle\pm|d_{1\downarrow}d_{1\uparrow}d_{2\downarrow}d_{2\uparrow}|\pm\rangle_{4e}=\pm \frac{\Gamma^*_{4e}}{2\sqrt{4(\epsilon-\epsilon_{\rm Q})^2+|\Gamma_{4e}|^2}}.
\end{equation}
It then follows the result Eq.~\eqref{Eq:Q-effective-model}.

\section{Noise}
\label{App:Noise}

Here, we briefly provide the details for the calculation of the auto- and cross-correlations, and we refer to Ref.~\cite{flindt2010} for further details. The out-of-equilibrium probabilities are obtained by solving the master equation
\begin{equation}
\sum_{a'}(w_{a\leftarrow a'}P_{a'}-w_{a'\leftarrow a}P_a)=0.
\end{equation}
We introduce the generalized matrix of rates $W(\boldsymbol{\chi})$, whose matrix elements can be computed by the Fermi golden rule rates 
\begin{equation}
    w(\boldsymbol{\chi})_{a \leftarrow a'}=\sum_{j\sigma}w^{j\sigma,{\rm in}}_{a\leftarrow a'}e^{-i\chi_j}+w^{j\sigma,{\rm out}}_{a\leftarrow a'}e^{i\chi_j},
\end{equation}
that feature the counting fields $\chi_j$. This way, defining $W\equiv W(\boldsymbol{\chi}=0)$, the master equation reads
\begin{equation}
    W\cdot {\bf P}=0,
\end{equation}
with ${\bf P}$ the vector of probabilities $P_a$ satisfying $\sum_aP_a=1$. The matrix $W(\boldsymbol{\chi})$ can be written in the form
\begin{equation}
    W(\boldsymbol{\chi})=W_{\rm diag}+\sum_j W^{\rm in}_je^{-i\chi_j}+W^{\rm out}_je^{i\chi_j},
\end{equation}
where the matrices $W_j^{\rm in/out}$ are purely off-diagonal in the sequential tunneling regime we are considering and the $\det W=0$ condition is enforced by the matrix diagonal matrix $W_{\rm diag}$, whose entries are the negative sum of the columns of the $W^{\rm in}_j$ and $W^{\rm out}_j$. Introducing the left eigenvector $\tilde{\bf P}$ satisfying $\tilde{\bf P}\cdot W=0$ and normalized such that $\tilde{\bf P}\cdot{\bf P}=1$, and defining
\begin{equation}
    W^{(n_i,n_j)}=\left.\frac{\partial^{(n_i+n_j)} W(\boldsymbol{\chi})}{\partial(i\chi_i)^{n_i}\partial (i\chi_j)^{n_j}}\right|_{\boldsymbol{\chi}=0}
\end{equation}
we can write the current as
\begin{equation}
    I_1=e\tilde{\bf P}\cdot W^{(1,0)}\cdot{\bf P},\qquad  I_2=e\tilde{\bf P}\cdot W^{(0,1)}\cdot{\bf P}
\end{equation}
and the current-current correlations as
\begin{eqnarray}
    S_{11}&=&e^2\tilde{\bf P}\cdot\left[W^{(2,0)}-2W^{(1,0)}RW^{(1,0)}\right]\cdot{\bf P}\\
    S_{22}&=&e^2\tilde{\bf P}\cdot\left[W^{(0,2)}-2W^{(0,1)}RW^{(0,1)}\right]\cdot{\bf P}\\
    S_{12}&=&e^2\tilde{\bf P}\cdot\left[W^{(1,1)}-W^{(1,0)}RW^{(0,1)}\right.\nonumber\\
    &-&\left.W^{(0,1)}RW^{(1,0)}\right]\cdot{\bf P},
\end{eqnarray}
with $R$ the pseudo inverse of $W$. For high bias voltage we have
\begin{eqnarray}
    S_{11}&=&e^2\tilde{\bf P}\cdot\left[W^{\rm in}_1-2W^{\rm in}_1RW^{\rm in}_1\right]\cdot{\bf P}\\
    S_{22}&=&e^2\tilde{\bf P}\cdot\left[W^{\rm in}_2-2W^{\rm in}_2RW^{\rm in}_2\right]\cdot{\bf P}\\
    S_{12}&=&-e^2\tilde{\bf P}\cdot\left[W^{\rm in}_1RW^{\rm in}_2\right.\nonumber\\
    &+&\left.W^{\rm in}_2RW^{\rm in}_1\right]\cdot{\bf P},
\end{eqnarray}
In the symmetric configuration we are considering we have
\begin{equation}
    I_1=e\tilde{\bf P}\cdot\left[W^{\rm in}_1\right]\cdot{\bf P}=e\tilde{\bf P}\cdot\left[W^{\rm in}_2\right]\cdot{\bf P}=I_2
\end{equation}
and the Fano factor is
\begin{eqnarray}
    F&=&1-\frac{\tilde{\bf P}\cdot\left[2W^{\rm in}_1RW^{\rm in}_1\right]\cdot{\bf P}}{\tilde{\bf P}\cdot\left[W^{\rm in}_1\right]\cdot{\bf P}}\nonumber\\
    &-&
    \frac{\tilde{\bf P}\cdot\left[W^{\rm in}_1RW^{\rm in}_2+W^{\rm in}_2RW^{\rm in}_1\right]\cdot{\bf P}}{\tilde{\bf P}\cdot\left[W^{\rm in}_1\right]\cdot{\bf P}}.
\end{eqnarray}

\section{Rates}
\label{App:rates}

We now provide explicit expressions for the rates in the case of one superconducting lead and for the two possibilities $\gamma_S'=0$ and $\gamma_S'\neq0$.  For $\gamma_S'=0$ the eigenstates can be written as
\begin{eqnarray}
    |\psi^e_a\rangle&=&C_{a,0}|0\rangle+\sum_{i=1,2}C_{a,di}|\!\uparrow\downarrow\rangle_i+C_{a,4e}|4e\rangle\nonumber\\
    |p\rangle_{j\sigma }&=&pv_{2e,\bar{p}}|\sigma\rangle_j-v_{2e,p}e^{-i\theta_{2e}}|\sigma t\rangle_j,
\end{eqnarray}
where the states $|S\rangle$, $|T0\rangle$, and $|T\sigma\rangle$ do not couple to all other states. Assuming very large positive bias voltage in the normal leads, the rate for an electron to tunnel in the DQD system reads
\begin{eqnarray}
    w^{\rm in}_{jp\sigma, a}&=&\sum_{is}e^{-i\chi_i}|{}_{j\sigma}\langle p|d^\dag_{is}|\psi_a^e\rangle|^2\\
    &=&\sum_{is}e^{-i\chi_i}|{}_{j\sigma}\langle p|\left(C_{a,0}|s\rangle_i+C_{a,d\bar{i}}|st\rangle_i\right)|^2 \\
    &=&\sum_{is}e^{-i\chi_i}\delta_{j,i}\delta_{\sigma,s}\left|pv^*_{2e,\bar{p}}C_{a,0}-v^*_{2e,p}e^{i\theta_{2e}}C_{a,d\bar{i}}\right|^2,\nonumber\\
    &=&e^{-i\chi_j}\left|pv^*_{2e,\bar{p}}C_{a,0}-v^*_{2e,p}e^{i\theta_{2e}}C_{a,d\bar{j}}\right|^2.
\end{eqnarray}
It then follows that the matrices $W^{\rm in}_1$ and $W^{\rm in}_2$ are in general different and can at most have the same values in different entries, such as the case for which $C_{a,d1}=C_{a,d2}$ occurring when the dots and the normal leads are in a symmetric configuration. 

In the case $\gamma_S'\neq 0$, the relevant basis states for the even and odd parity sector
\begin{eqnarray}
    |\psi^e_a\rangle&=&C_{a,0}|0\rangle+\sum_{i=1,2}C_{a,di}|\!\uparrow\downarrow\rangle_i+C_{a,S}|S\rangle+C_{a,4e}|4e\rangle\nonumber\\
    |p\rangle_{\tau\sigma }&=&pv_{2e,\bar{p},\tau}|\sigma\rangle_\tau-e^{-i\theta_{2e,\tau}}v_{2e,p,\tau}|\sigma t\rangle_\tau,
\end{eqnarray}
The general matrix element reads
\begin{eqnarray}
    {}_{\tau\sigma}\langle p|d^\dag_{is}|\psi_a^e\rangle&=&\frac{\delta_{\sigma s}}{\sqrt 2}(\delta_{i,1}+\tau\delta_{i,2})\nonumber\\
    &\times& (pv^*_{2e,\bar{p}}C_{a,0}- v^*_{2e,p}e^{i\theta_{2e}}C_{a,d\bar{i}}-\frac{\tau}{\sqrt 2}C_{a,S}).\nonumber\\
\end{eqnarray}
For very large positive bias voltage in the normal leads, the rate for an electron to tunnel in the DQD system now reads
\begin{widetext}
\begin{eqnarray}
    w^{\rm in}_{\tau p\sigma, a}&=&\sum_{is}e^{-i\chi_i}\left|{}_{\tau\sigma}\langle p|d^\dag_{is}|\psi_a^e\rangle\right|^2\\
    &=&\sum_{is}e^{-i\chi_i}\left|{}_{\tau\sigma}\langle p|\left(C_{a,0}|s\rangle_i+C_{a,d\bar{i}}|st\rangle_i+C_{a,S}d^\dag_{is}|S\rangle\right)\right|^2 \nonumber\\
    &=&\sum_{is}e^{-i\chi_i}\left|\frac{\delta_{\sigma s}}{\sqrt 2}(\delta_{i,1}+\tau\delta_{i,2})(pv^*_{2e,\bar{p}}C_{a,0}- v^*_{2e,p}e^{i\theta_{2e}}C_{a,d\bar{i}}-\frac{\tau}{\sqrt 2}C_{a,S})\right|^2,\nonumber\\
    &=&\frac{1}{2}e^{-i\chi_1}\left|(pv^*_{2e,\bar{p},\tau}C_{a,0}- v^*_{2e,p,\tau}e^{i\theta_{2e}}C_{a,d2}-\frac{\tau}{\sqrt 2}C_{a,S})\right|^2\nonumber\\
    &+&\frac{1}{2}e^{-i\chi_2}\left|(pv^*_{2e,\bar{p},\tau}C_{a,0}- v^*_{2e,p,\tau}e^{i\theta_{2e}}C_{a,d1}-\frac{\tau}{\sqrt 2}C_{a,S})\right|^2
\end{eqnarray}
Now, if by symmetry we have $C_{a,d1}=C_{a,d2}=C_{a,d}$, the rate becomes 
\begin{equation}
    w^{\rm in}_{\tau p\sigma, a}=\frac{1}{2}(e^{-i\chi_1}+e^{-i\chi_2})\left|(pv^*_{2e,\bar{p},\tau}C_{a,0}- v^*_{2e,p,\tau}e^{i\theta_{2e}}C_{a,d}-\frac{\tau}{\sqrt 2}C_{a,S})\right|^2
\end{equation}
\end{widetext}

\section{Josephson current}

At the microscopic level, the superconducting leads $S_i$, with $i=0,1,2$, are connected to the DQD system via tunneling Hamiltonian terms of the type of Eq.~\eqref{Eq:Htun}. By employing the Meir-Wingreen formula \cite{meir1992}, we obtain an expression of the current that is exact at all orders in $\gamma_S,\gamma_S'$. Furthermore, following Ref.~\cite{pala2007,governale2008}, in the limit $|\Delta|\to \infty$ the dissipationless Josephson current in the superconducting lead $S_\eta$, with $\eta=0,1,2$ reads
\begin{equation}\label{Eq:Josephson}
I^J_\eta=\frac{e}{\hbar}{\rm Re}{\rm Tr}\left[\tau_3\boldsymbol{\Gamma}^a_\eta \int \frac{d\omega}{2\pi}{\bf G}^<(\omega)\right],
\end{equation}
where the third Pauli matrix $\tau_3$ and the Fourier transform ${\bf G}^<(\omega)$  of the exact lesser DQD Green's function ${\bf G}^<_{nm}(t)\equiv i\langle \mathbf{\Phi}_n^\dag(0)\mathbf{\Phi}_m(t)\rangle$ are defined in the Nambu space with $\mathbf{\Phi}=(d_{1\uparrow},d_{2,\uparrow},d^\dag_{1\downarrow},d^\dag_{2\downarrow})^T$. Furthermore, we have defined $\boldsymbol{\Gamma}_\eta^a=\Gamma_\eta \otimes(e^{i\phi_\eta}\tau_++e^{-i\phi_\eta}\tau^-)$, with $\phi_\eta$ the phase of $S_\eta$, $\tau^\pm$ raising and lowering matrices in the Nambu space, and
\begin{eqnarray}
\Gamma_\eta&=&\left(\begin{array}{cc}
\gamma_{S,\eta} &  \gamma'_{S,\eta}  \\
\gamma'_{S,\eta}   & \gamma_{S,\eta}
\end{array}\right).
\end{eqnarray}
Having chosen the terminal $S_0$ as a phase reference in the setup of Fig.~\ref{fig-outline}(a), the terminals $S_1$ and $S_2$ are only coupled locally to the nearby QD, i.e. $\gamma'_{S,1}=\gamma'_{S,2}=0$.
The integral over the frequency of the DQD Green's function amounts to the equal time lesser Green's function $\int \frac{d\omega}{2\pi}{\bf G}^<_{nm}(\omega)=i\langle \Phi_n^\dag\Phi_m \rangle$, and the expectation value is taken on the DQD density matrix, so that the current reads $I^J_\eta=-\frac{2e}{\hbar}{\rm Im Tr}\left[\Gamma_\eta e^{i\phi_\eta}F\right]$,
with $F_{ij}=\langle d_{i\downarrow}d_{j\uparrow}\rangle$ the equal time anomalous Green's function computed over the out-of-equilibrium DQD density matrix. 

Here we provides details concerning the Josephson current. The latter, in the lead $S_i$ with $i=1,2$ can be written as
\begin{eqnarray}
    I^J_i&=&-(2e\gamma_S/\hbar){\rm Im}\left[e^{i\varphi_i}\langle d_{i\downarrow}d_{i\uparrow}\rangle\right]\\
    &=&-(2e\gamma_S/\hbar)\sum_{a}P_a{\rm Im}\left[e^{i\varphi_i} \langle\psi_a|d_{i\downarrow}d_{i\uparrow}|\psi_a\rangle\right].
\end{eqnarray}
Since the superconducting leads $S_i$ is  tunnel-coupled to only dot $i$, for $i=1,2$, no contribution proportional to $\gamma_S'$ arises in the current.

\bibliography{quartets-bib}{}

\end{document}